\pdfoutput=1

\newcommand*{\ATLASLATEXPATH}{}

\documentclass[USenglish,texlive=2016,cernpreprint]{atlasdoc}

\usepackage{atlaspackage}
\usepackage{siunitx}

\usepackage{atlasbiblatex}

\usepackage{atlascontribute}

\usepackage{atlasphysics}
\usepackage{atlasbsm}

\usepackage{EXOT-2017-03-PAPER-defs}
\usepackage{multirow}

\AtlasTitle{Search for long-lived particles in final states with displaced dimuon vertices in $pp$ collisions at $\sqrt{s}=13~\tev$ with the ATLAS detector}

\author{The ATLAS Collaboration}

\AtlasRefCode{EXOT-2017-03}

\AtlasNote{EXOT-2017-03}

\PreprintIdNumber{CERN-EP-2018-173}

\AtlasDate{\today}

\AtlasJournalRef{Phys.\ Rev.\ D 99 (2019) 012001}
\AtlasDOI{10.1103/PhysRevD.99.012001}

\AtlasAbstract{%
A search is performed for a long-lived particle decaying into a final state that includes a pair of muons of opposite-sign electric charge, using proton--proton collision data collected at $\sqrt{s}=13~\tev$ by the ATLAS detector at the Large Hadron Collider corresponding to an integrated luminosity of $\lumi$.
No significant excess over the Standard Model expectation is observed.  Limits at  $95\%$ confidence level on the lifetime of the long-lived particle are presented in models of new phenomena including gauge-mediated supersymmetry or decay of the Higgs boson, $\Hboson$, to a pair of dark photons, $\zd$. Lifetimes in the range c$\tau = 1$--$2400~\cm$ are excluded, depending on the parameters of the model. In the supersymmetric model, the lightest neutralino is the next-to-lightest supersymmetric particle, with a relatively long lifetime due to its weak coupling to the gravitino, the lightest supersymmetric particle.  The lifetime limits are determined for very light gravitino mass and various assumptions for the neutralino mass in the range $300$~\GeV\ to $1000~\gev$.  In the dark photon model, the lifetime limits are interpreted as exclusion contours in the plane of the coupling between the $\zd$ and the Standard Model $\Zboson$ boson versus the $\zd$ mass (in the range $20$--$60~\gev$), for various assumptions for the $H\to \zd \zd$ branching fraction.
}

\hypersetup{pdftitle={ATLAS document},pdfauthor={The ATLAS Collaboration}}

\begin{document}

\maketitle

\section{Introduction}
\label{sec:intro}

The ATLAS and CMS experiments at the Large Hadron Collider (LHC) were conceived to address a variety of questions not fully explained within the Standard Model (SM) of particle physics.  
The data collected by the LHC experiments have not yet revealed evidence of physics beyond the Standard Model (BSM).  As a result, there is an increased emphasis on the exploration of unusual final-state signatures that would elude the searches based on experimental methods aimed at prompt signatures.  In many models of BSM physics there are free parameters that influence the lifetimes of the new particle states, with no strong motivation for assuming that all the particles decay promptly\footnote{For the purposes of this analysis, a promptly decaying particle is one with a lifetime no larger than a few tens of picoseconds.} and thus give final states investigated with standard analysis techniques. Nor are there any strong demands that these are stable on the detector scale and only weakly interacting, leading to missing transverse momentum signatures. Particle lifetimes in the SM, for instance, span roughly $28$ orders of magnitude~\cite{Patrignani:2016xqp}, from the strong decay to the scale of the neutron lifetime. There are a number of BSM models where long-lived particles (LLPs) arise naturally~\cite{FAIRBAIRN20071,ZD1-1312-4992}.  Supersymmetry (SUSY)~\cite{Golfand:1971iw,Volkov:1973ix,Wess:1974tw,Wess:1974jb,Ferrara:1974pu,Salam:1974ig} with $R$-parity violation~\cite{Barbier:2004ez,Allanach:2006st}, general gauge-mediated (GGM) supersymmetry breaking~\cite{Dimopoulos:1996vz,Cheung:2007es,Meade:2008wd} and split supersymmetry~\cite{Hewett:2004nw,Arvanitaki:2012ps} are examples where small couplings, mass scales associated with the BSM physics or heavy mediator particles, respectively, lead to high-mass (greater than a few hundred $\gev$) LLPs.  Scenarios with low-mass LLPs include hidden-valley models~\cite{Han:2007ae}, stealth supersymmetry~\cite{Fan:2011yu} and dark-sector gauge bosons~\cite{Schuster:2009au,ZD1-1312-4992}.

Events with long-lived particles may feature vertices that are significantly displaced from the proton--proton ($pp$) interaction point (IP).  
This article presents the results of a search for displaced vertices (DVs) formed by a pair of muons of opposite-sign electric charge, denoted ``OS'' muons.  The search is designed to be sensitive to decays of LLPs with masses between $20~\gev$ and $1100~\gev$ and DVs at distances ranging from a few centimeters to a few meters from the IP.  The data sample consists of $pp$ collisions at $\sqrt{s} = 13~\tev$ and an integrated luminosity of \lumi\ collected with the ATLAS detector at the LHC.

Although SM decay products typically consist primarily of hadrons, due to the relatively large number of color degrees of freedom for quarks, there are notable advantages to searching for DVs using only tracks of identified muons:  the design of the ATLAS muon spectrometer allows detection of dimuon DVs within an unusually large decay volume, free from backgrounds associated with vertices produced in interactions of hadrons with detector material~\cite{Aaboud:2016poq,Aaboud:2017pjd}.

Previous searches by the ATLAS Collaboration for high-mass LLPs that decay within the inner detector to give displaced dilepton vertices excluded LLP lifetimes of $c\tau = 0.1$--$100~\cm$~\cite{SUSY-2014-02}.  ATLAS has also searched for very low mass LLPs ($< 10~\gev$) by considering pairs of highly collimated leptons \cite{EXOT-2013-22}, with sensitivity to LLP lifetimes of  $c\tau = 0.1$--$20~\cm$.
Several other LLP searches targeting a wide range of lifetimes and signatures have been conducted by the ATLAS~\cite{EXOT-2011-12, SUSY-2011-02, SUSY-2012-14, SUSY-2012-15, EXOT-2012-28, SUSY-2013-01, SUSY-2013-17, EXOT-2013-12, SUSY-2016-02, SUSY-2016-03}, CMS~\cite{CMS-EXO-11-101, CMS-EXO-11-035, CMS-B2G-12-024, CMS-EXO-12-037, CMS-EXO-12-038, CMS-SUS-14-020, CMS-EXO-15-010}, LHCb~\cite{LHCb_1412.3021, LHCb_1508.04094, LHCb_1612.00945, LHCb_1612.07818}, CDF~\cite{CDF_9805017}, D0~\cite{D0_0607028, D0_0906.1787}, BaBar~\cite{BaBar_1502.02580}, Belle~\cite{Belle_1301.1105}, and ALEPH~\cite{ALEPH_0203024} collaborations.

\section{ATLAS detector}
\label{sec:detector}

The ATLAS detector~\cite{PERF-2007-01,ATLAS-TDR-19} at the LHC covers nearly the entire solid angle around the collision point.\footnote{ATLAS uses a right-handed coordinate system with its origin at the nominal interaction point in the center of the detector and the $z$-axis along the beam pipe.  The $x$-axis points from the IP to the center of the LHC ring, and the $y$-axis points upwards.  Cylindrical coordinates $\left(r,\phi\right)$ are used in the transverse plane, $\phi$ being the azimuthal angle around the $z$-axis.  The pseudorapidity is defined in terms of the polar angle $\theta$ as $\eta = -\ln \, \tan\left(\theta/2\right)$.  Angular distances are measured in units of $\Delta R \equiv \sqrt{\left(\Delta \eta\right)^2 + \left(\Delta \phi\right)^2}$.}
It consists of an inner tracking detector surrounded by a thin superconducting solenoid, electromagnetic and hadronic calorimeters,
and a muon spectrometer incorporating superconducting toroidal magnets.

The inner detector (ID) is immersed in a \SI{2}{\tesla} axial magnetic field 
and provides charged-particle tracking in the range $|\eta| < 2.5$.
A high-granularity silicon pixel detector covers the vertex region and typically provides four measurements per track, 
the first hit being normally in the innermost layer.
It is followed by a silicon microstrip tracker, which usually provides four two-dimensional measurement points per track.
These silicon detectors are complemented by a transition radiation tracker,
which enables radially extended track reconstruction up to $|\eta| = 2.0$. 
The transition radiation tracker also provides electron identification information 
based on the fraction of hits (typically 30 in total) above a higher energy-deposit threshold corresponding to transition radiation.

The calorimeter system covers the pseudorapidity range $|\eta| < 4.9$.
Within the region $|\eta|< 3.2$, electromagnetic calorimetry is provided by barrel and 
endcap high-granularity lead/liquid-argon (LAr) sampling calorimeters,
with an additional thin LAr presampler covering $|\eta| < 1.8$ to correct for energy loss in material upstream of the calorimeters.
Hadronic calorimetry is provided by a steel/scintillator-tile calorimeter,
segmented into three barrel structures within $|\eta| < 1.7$, and two copper/LAr hadronic endcap calorimeters.
The solid-angle coverage is completed with forward copper/LAr and tungsten/LAr calorimeter modules
optimized for electromagnetic and hadronic measurements respectively.

The muon spectrometer (MS) comprises separate trigger and high-precision tracking chambers measuring the deflection of muons in a magnetic field generated by three superconducting air-core toroidal magnets, each with eight coils. The field integral of the toroids ranges between \num{2.0} and \SI{6.0}{\tesla\metre}  
across most of the detector.  The MS is designed to detect muons in the region $|\eta| < 2.7$ and to provide momentum measurements with a relative resolution better than $3\%$ over a wide transverse momentum ($\pt$) range and up to $10\%$ at $\pt \sim 1~\tev$.  It consists of a barrel ($|\eta| < 1.05$), with an inner radius of about $500~\cm$, and two endcap sections ($1.05 < |\eta| < 2.7$).

Resistive-plate chambers in the barrel and thin-gap chambers in the endcap regions provide triggering capability to the detector as well as $\left(\eta,\phi\right)$ position measurements with a typical spatial resolution of $5$--$10~\mm$.  A precise momentum measurement is provided by three layers of monitored drift-tube chambers (MDT), with each chamber providing six to eight $\eta$ measurements along the muon trajectory.  For $|\eta| > 2$, the inner layer is instrumented with a quadruplet of cathode-strip chambers (CSC) instead of MDTs.  The single-hit resolution in the bending plane for the MDT and the CSC is about $80~\mu$m and $60~\mu$m, respectively.  The muon chambers are aligned with a precision between $30~\mu$m and $60~\mu$m.  The material between the IP and the MS ranges from approximately $100$ to $190$ radiation lengths, depending on $\eta$, and consists mostly of the calorimeters.

Online event selection is performed with a two-level trigger system~\cite{Aaboud:2016leb}.  A hardware-based level-1 trigger which uses information from the MS trigger chambers and the calorimeters is followed by a software-based trigger.

\section{Data and simulated samples}
\label{sec:mcsamples}

Proton-proton collision data, collected at the LHC during 2016, with a center-of-mass energy $\sqrt{s}=13~\tev$, are analyzed.  After application of detector and data-quality requirements, the integrated luminosity of the data sample is \lumi.  

Samples of Monte Carlo (MC) simulated events are used for studies of both the LLP signal and background processes.  The detector response was simulated with \textsc{Geant4} \cite{Agostinelli:2002hh,Aad:2010ah}, and the events were processed with the same reconstruction software as used for the data.  The distribution of the number of additional $pp$ collisions in the same or neighboring bunch crossings (``pileup'') is accounted for by overlaying minimum-bias events simulated with \textsc{Pythia8}~\cite{Sjostrand:2014zea} using the A2 set of tuned parameters (tune)~\cite{ATL-PHYS-PUB-2011-014} and MSTW2008LO parton distribution function (PDF) set~\cite{Sherstnev:2007nd}. The pileup profile in the MC samples is reweighted to match the distribution observed in the data.  

\subsection{BSM signal samples}
\label{sec:signalMC}

\begin{figure}[!htb]
\centering
\subfloat[]{\includegraphics[width=0.45\textwidth]{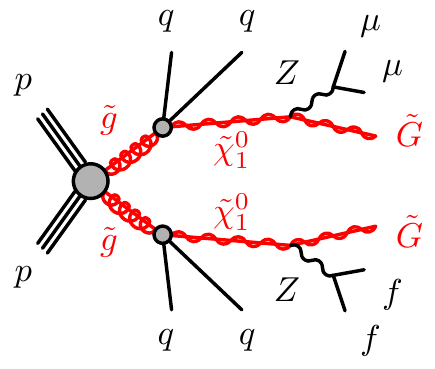}} \hspace{0.08\textwidth}
\subfloat[]{\includegraphics[width=0.45\textwidth]{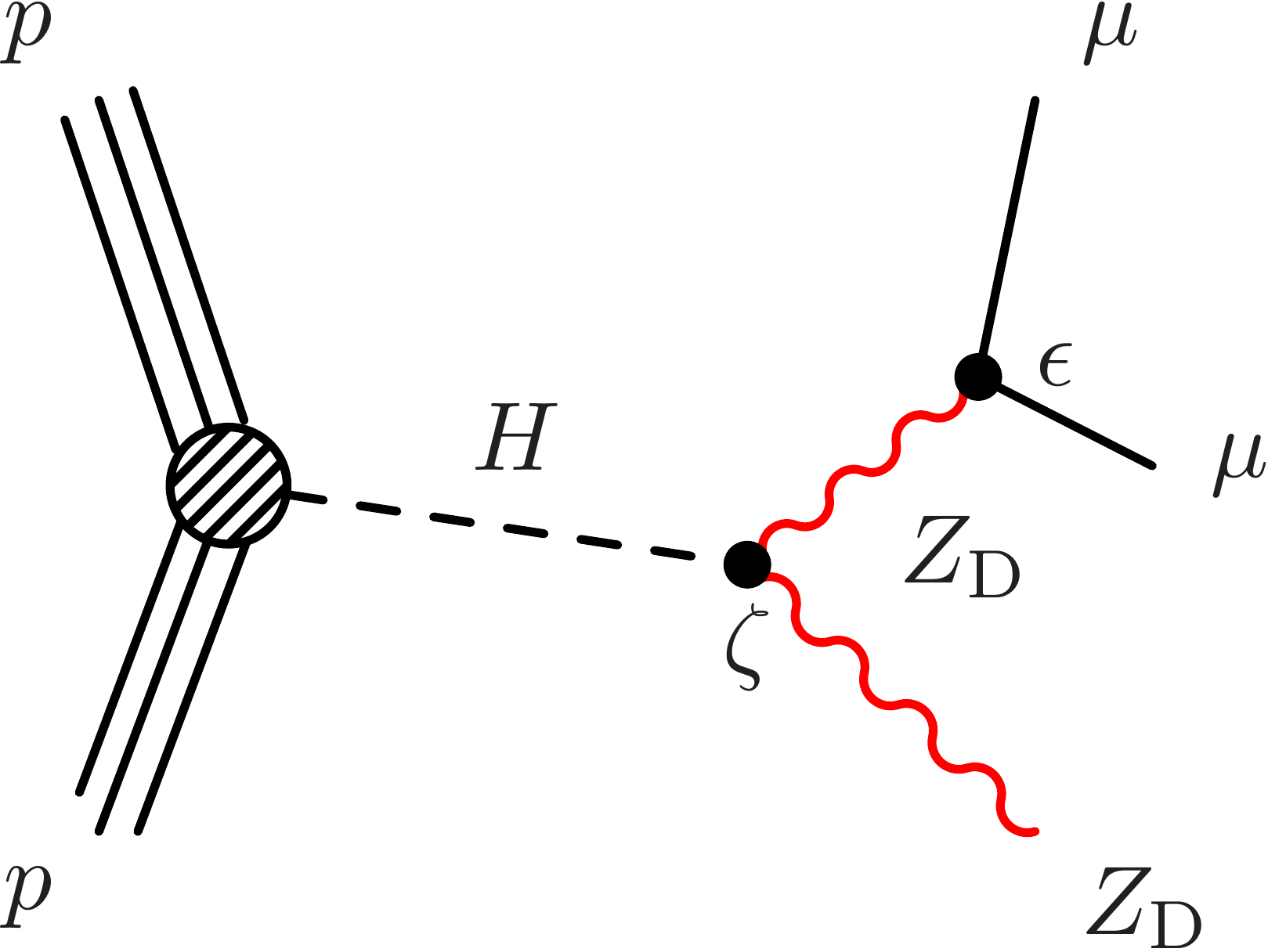}}
\caption{Diagrams representing BSM processes considered signals in this article:  (a)  long-lived neutralino $\ninoone$ decay in a GGM scenario, and (b) long-lived dark photons $\zd$ produced from Higgs boson decay.  The quarks, $q$, may have different flavors (excluding the top quark).  The symbol $f$ represents fermions lighter than half the mass of the $\Zboson$ boson.
\label{fig:feynsignals} }
\end{figure}

Monte Carlo simulated samples from two different BSM physics models are used to tune selection criteria and to evaluate signal efficiencies for use in converting signal yields into cross-sections.  The chosen models, a general gauge-mediated supersymmetry and dark-sector gauge boson model, represent a variety of BSM physics possibilities, as well as final-state topologies and kinematics, to which the analysis may be sensitive.  The two processes are illustrated  in \Fig{\ref{fig:feynsignals}}.  Samples for both models were generated with MadGraph5$\_$aMC@NLO \cite{Alwall:2014hca} using the NNPDF23LO PDF set \cite{Martin:2009iq} and \textsc{Pythia8} for parton showering and hadronization.  The matrix elements were calculated to next-to-leading order in the strong coupling constant.
The EvtGen generator \cite{Lange:2001uf} was used for weak decays of heavy-flavor hadrons.  The hadronization and underlying-event parameters were set according to the A14 tune~\cite{ATL-PHYS-PUB-2011-014}.

In $R$-parity-conserving (RPC) SUSY models where gauge interactions mediate the breaking of the supersymmetry, the gravitino $\tilde{G}$ acquires its mass from a ``super-Higgs'' mechanism and may be very light:  $m_{\gravino} = {\cal O}\left(\kev\right)$. The mass is given by:

\begin{equation}
m_{\gravino} = \frac{F_0}{\sqrt{3}M_\text{Pl}} = \left(\frac{\sqrt{F_0}}{100~\tev}\right)^2\times 2.4~\ev\, ,
\label{eqn:mgravitino}
\end{equation}

where $\sqrt{F_0}$ is the fundamental scale of supersymmetry breaking, typically $\gtrsim 100~\tev$, and $M_\text{Pl}$ is the Planck scale.  Hence, the gravitino is the lightest supersymmetric particle (LSP).  All heavier supersymmetric particles decay promptly through cascades leading to the next-to-lightest supersymmetric particle (NLSP), which then decays into the LSP gravitino via an interaction with a $1/F_0$ suppression.  The NLSP, depending on model choices, is either the lightest slepton or lightest neutralino, $\ninoone$.  For the latter case, chosen for this search and described in Ref.~\cite{GGM-1006-4575}, if $\ninoone$ has a significant wino or higgsino component the branching fraction for the decay $\ninoone \to \Zboson\gravino$ can be ${\cal O}\left(1\right)$.  
The lifetime of the $\ninoone$ is determined by $F_0$ (or, alternatively, by $m_{\gravino}$, according to \Eqn{(\ref{eqn:mgravitino})}) and its mass $m_{\ninoone}$,

\begin{equation*}
c\tau_{\ninoone} \propto \frac{16\pi F_0^2}{m_{\ninoone}^5} \approx \left(\frac{100~\gev}{m_{\ninoone}}\right)^5 \left(\frac{\sqrt{F_0}}{300~\tev}\right)^4 \times 1~\cm \,,
\end{equation*}
and hence $\ninoone$ is long-lived (i.e.\ non-prompt) for $\sqrt{F_0} = 10^3~\tev$ to $10^4~\tev$.

In the GGM model, a $pp$ interaction creates a pair of gluinos, followed by a cascade of decays leading to $\ninoone \to Z\gravino$.  A simplified model is used whereby the cascade of decays of SUSY particles is reduced to a single vertex:  $\gluino\to q q\ninoone$, where $q$ represents any of the quarks lighter than the top quark, with equal probability for each.  Six signal samples were generated with $m_{\gluino}=1.1~\tev$ and $\ninoone$ masses and lifetimes given in \Tab{\ref{tab:ggmmc}}.  
The value of $1.1~\tev$ for the gluino mass was chosen to be consistent with the value used in Ref.~\cite{SUSY-2014-02}, the previous search for DVs with a GGM interpretation.
The signal cross-sections are calculated to next-to-leading order in the strong coupling constant, adding the resummation of soft gluon emission at next-to-leading-logarithm accuracy (NLO+NLL)~\cite{Beenakker:1996ch,Kulesza:2008jb,Kulesza:2009kq,Beenakker:2009ha,Beenakker:2011fu}.  The nominal cross-sections and their uncertainties are taken from an envelope of cross-section predictions using different PDF sets and factorization and renormalization scales, as described in Ref.~\cite{Borschensky:2014cia}.

\begin{table}[htb]
\centering
\caption{ \label{tab:ggmmc}
 MC signal samples for the GGM SUSY interpretation. For a given $m_{\ninoone}$, the gravitino mass is chosen to give the desired lifetime. For all samples, $m_{\gluino}=1100~\gev$, $\sigma\left(pp\to \gluino\gluino \right) = 163.5~\fb$, $B\left(\ninoone \to \Zboson \gravino \right) = 1.0$ and $B\left(\Zboson \to \mu^+\mu^- \right) = 0.03366$.}
 \begin{tabular}{rr}
\toprule
$m_{\ninoone}~[\gev]$ & $c\tau_{\ninoone}~[\cm]$ \\[+6pt]
\midrule
$ 300$ & $100$ \\
$ 300$ & $500$ \\
$ 700$ & $100$ \\
$ 700$ & $500$ \\
$1000$ & $100$ \\
$1000$ & $500$ \\
\bottomrule
 \end{tabular}
 
\end{table}

A number of BSM theories feature a ``hidden'' or ``dark'' sector of matter that does not interact directly with SM particles but may nevertheless interact weakly with SM matter via coupling to the Higgs field.  These are ``Higgs portal'' models that address the dark-matter problem and electroweak baryogenesis.  The model considered for this search is one in which there exists a $U(1)_\text{D}$ symmetry in the dark sector, and the dark vector gauge boson $\zd$, often called a ``dark photon,'' is given mass via a singlet scalar field $\Hd$ that breaks the symmetry and is analogous to the Higgs field $\Hboson$ in the visible SM sector~\cite{ZD1-1312-4992,ZD2-1412-0018}.  

The BSM terms in the Lagrangian density include both a hypercharge portal and a Higgs portal, providing kinetic $\Zboson$--$\zd$ mixing (i.e.\ mixing between $U(1)_Y$ and $U(1)_\text{D}$) and $\Hboson$--$\Hd$ mixing, regulated by the small coupling parameters $\epsilon$ and $\zeta$, respectively.  There are two vector-boson mass eigenstates, one that is mostly $\zd$ and another that is mostly SM $\Zboson$, as well as two scalar mass eigenstates, one that is mostly $\Hd$ and another that is mostly $\Hboson$.  For simplicity, the physical (mass) states are denoted by $\Hboson$, $\Hd$, $\Zboson$ and $\zd$.

In the scenario where the singlet scalar $\Hd$ is heavier than the SM $\Hboson$ boson, which means that the process $\Hboson \to \Hd\Hd$ is kinematically forbidden, and $\zd$ is lighter than half the $\Hboson$ mass, events with a displaced dimuon vertex signature would be observable in experiments at the LHC.  The $\zd$ bosons are produced on-shell in Higgs boson decays and decay to SM fermions due to their induced couplings to the electroweak current.  A small value of $\epsilon$ ($\lesssim 10^{-5}$) results in a long-lived $\zd$ state:  $c\tau_{\zd} \propto 1/\epsilon^{2}$.
The branching fraction for $\Hboson \to \zd\zd$ is determined by the value of $\zeta$ and the masses of the scalar singlets:

\begin{equation*}
B\left(\Hboson \to \zd\zd\right) \propto \zeta {\frac{m^2_H}{| m^2_{\Hd}-m^2_H |}}\, ,
\end{equation*}
where values as large as $25\%$ have not yet been ruled out by constraints from Higgs coupling fits~\cite{HIGG-2015-03,CMS-HIG-16-016}.  For $\epsilon \ll 1$, the $\zd$ branching fraction to muons, $B\left(\zd\to \mu^+\mu^-\right)$, is independent of $\epsilon$ but varies with $\mzd$~\cite{ZD2-1412-0018}:  from a value of $0.1475$ for $\mzd = 20~\gev$ to a value of $0.1066$ for $\mzd = 60~\gev$.  Five signal samples were generated with $\zd$ masses and lifetimes given in \Tab{\ref{tab:darkzmc}}.  The Higgs boson is produced via the gluon-gluon fusion process, assuming a cross-section of $44.1~\pb$, calculated at next-to-next-to-leading order in the strong coupling constant, adding the resummation of soft gluon emission at next-to-next-to-leading-logarithmic accuracy~\cite{deFlorian:2016spz}.  The inclusion of other production processes was found to have a negligible impact on the analysis.

\begin{table}[htb]
\centering
 \caption{\label{tab:darkzmc}
 MC signal samples for the dark-sector interpretation.  For all samples, $m_{\Hboson}=125~\gev$, $m_{\Hd}=300~\gev$, $\sigma\left(pp\to \Hboson \right) = 44.1~\pb$ (via the gluon-gluon fusion production process) and $B\left(\Hboson \to \zd \zd \right) = 0.10$.}
 \begin{tabular}{rrr}
\toprule
 $m_{\zd}~[\gev]$ & $c\tau_{\zd}~[\cm]$ & $B\left(\zd\to \mu^+\mu^-\right)$  \\ [+6pt]
\midrule
$  20$ & $50$ & 0.1475 \\
$  40$ & $50$ & 0.1370 \\
$  40$ & $500$ & 0.1370  \\
$  60$ & $50$ & 0.1066 \\
$  60$ & $500$ & 0.1066 \\
\bottomrule
 \end{tabular}
 
\end{table}

The signal samples were generated with values of the LLP lifetime that were chosen to provide sufficiently large number of DVs across the full fiducial decay volume of the search:  approximately $0 < \rvtx < 400~\cm$.
To obtain distributions corresponding to a different lifetime, $c\tau_\text{new}$, each event is given a weight. The weight $w_i$ assigned to each LLP $i$ is computed as:

\begin{equation*}
w_i\left( t_i\right) =   \frac{\tau_\text{gen}}{\text{e}^{-t_i/\tau_\text{gen}}}\cdot \frac{\text{e}^{-t_i/\tau_\text{new}}}{\tau_\text{new}} \, ,
\end{equation*}

where the first factor reweights the exponential decay to a constant distribution and the second factor reweights to the desired lifetime.  The quantity $t_i$ is the proper decay time of the LLP and $c\tau_\text{gen}$ is the lifetime assumed in generating the sample.  The event-level weight is the product of the weights for the two LLPs in each event.  The event-level signal efficiency is then the sum of weights for all events for which at least one reconstructed dimuon vertex satisfies the selection criteria, divided by the total number of events generated.  This scheme ensures that any dependence of the efficiency on the decay time of both LLPs in the event, and not just the one decaying to a dimuon final state, is properly taken into account for each choice of $c\tau_\text{new}$.

The lifetime reweighting technique is validated by using a signal sample of a given $c\tau_\text{gen}$ to predict the efficiency for a different lifetime and comparing with the value directly obtained from a sample generated with that lifetime.

\subsection{SM background samples}
\label{sec:smsamples}

The MC simulations of background processes are used only as a guide for some of the selection criteria and for categorization of the types of background, while the background yield itself is predicted from techniques that use solely the data.  The MC generators, hadronization and showering software packages, underlying-event simulation and choice of parton distribution functions are summarized in \Tab{\ref{tab:mcgen}}. Further details about the generator settings used for these processes can also be found in Refs.~\cite{ATL-PHYS-PUB-2016-004,ATL-PHYS-PUB-2017-006,ATL-PHYS-PUB-2016-002,ATL-PHYS-PUB-2017-007,ATL-PHYS-PUB-2017-005}.

\begin{table}[htb]
\centering
\caption{The MC generators, hadronization and showering software package, underlying-event simulation and PDF sets used for the simulated background events. The mass-range of the low-mass Drell--Yan sample is restricted to $6 <\mvtx < 60~\gev$. \label{tab:mcgen}}
\resizebox{\linewidth}{!}{
 \begin{tabular}{lllll}
\toprule
Sample & MC generator & Hard-process PDF & Hadronization and showering & Non-perturbative tune and parton-shower PDF \\
\midrule

$\Zboson+\text{jets}$ &  \textsc{Powheg}~\cite{Alioli:2010xd,Alioli:2008gx} & CT10~\cite{Lai:2010vv} & \textsc{Pythia8}~\cite{Sjostrand:2014zea}+EvtGen~\cite{Lange:2001uf} &  AZNLO+CTEQ6L1~\cite{STDM-2012-23} \\
Low-mass Drell--Yan &  \textsc{Powheg} & & \textsc{Pythia8}+EvtGen & AZNLO+CTEQ6L1 \\
$\ttbar$ &  \textsc{Powheg} & CT10 & \textsc{Pythia6}~\cite{Sjostrand:2000wi}+EvtGen &  P2012~\cite{Skands:2010ak}+CTEQ6L1 \\
$\Wboson+\text{jets}$ &  \textsc{Powheg} & CT10 & \textsc{Pythia8} & AZNLO+CTEQ6L1   \\
$\Zboson\Zboson$ &  \textsc{Powheg-Box} v2~\cite{Nason:2013ydw} & CT10nlo & \textsc{Pythia8} & AZNLO+CTEQ6L1  \\
$\Wboson\Wboson$ & \textsc{Powheg-Box} v2 & CT10nlo & \textsc{Pythia8} & AZNLO+CTEQ6L1  \\
$\Wboson\Zboson$ &  \textsc{Powheg-Box} v2 & CT10nlo & \textsc{Pythia8} & AZNLO+CTEQ6L1  \\
Single top &  \textsc{Powheg}~\cite{Alioli:2009je,Frederix:2012dh} & CT10 & \textsc{Pythia6} & P2012+CTEQ6L1  \\

\bottomrule
 \end{tabular}
} 
\end{table}

Each of the simulated background samples is scaled to correspond to an integrated luminosity of $\lumi$, the size of the data sample.

\section{Event selection, signal efficiencies and background estimate}
\label{sec:eventsel}

Candidate signal events are selected by identifying $\mu^+\mu^-$ pairs consistent with having been produced in a vertex displaced at least several centimeters from the IP.\footnote{The RMS spread of the $z$ distribution of the $pp$ interaction vertices is $47.7~\mm$ and the spreads in the $x$ and $y$ directions are less than $0.01~\mm$.}  The selection criteria are designed to strongly suppress background from SM processes that produce muons near the IP while efficiently accepting signal events over a wide range of LLP masses, lifetimes and velocities.  To retain the greatest possible model independence, minimal requirements are placed on other aspects of the event.

The initial event selection is performed with a combination of triggers that require either the presence of a muon candidate or large missing transverse momentum, whose magnitude is denoted $\met$.  Next, offline selection criteria are used to first identify suitable muon candidates, and then pairs of muons of opposite charge consistent with a DV.  The backgrounds from all SM beam-collision and non-beam-collision processes (cosmic-ray muons or beam-halo particles) are estimated directly from the data.  Finally, the number of vertices expected from background processes is compared with the observed number of vertices in data in two signal regions, distinguished by the dimuon invariant mass.

\subsection{Trigger requirements}
\label{sec:triggers}

Events must satisfy the requirements of at least one of four different triggers in order to achieve the best possible efficiency for a wide variety of signal topologies and kinematics. The triggers used and their descriptions are given in \Tab{\ref{tab:trigDesc}}.  The first two triggers are highly efficient for signals with high-mass states that feature muons with large transverse momentum and large transverse impact parameters, $d_0$, such as the GGM model, while the final two allow efficient selection of signals featuring low-mass states, and therefore lower-$\pt$ muons (e.g.\ the dark-sector model).  All three of the muon triggers use only measurements in the MS to identify muons. 

\begin {table}[!htb]
\centering
\caption{
   \label{tab:trigDesc}
Description of triggers used to select events.  The quantity $\dr_{\mu\mu}$ is the angular distance between the two muons in the collimated-dimuon trigger.}
\resizebox{\linewidth}{!}{
\begin{tabular}{llll}
\toprule
Signal type & Trigger & Description & Thresholds \\

\midrule
\multirow{2}{*}{High mass} & \met & missing transverse momentum & $\met > 110~\gev$ \\
&single muon & single muon restricted to the barrel region & muon $|\eta | < 1.05$ and $\pt > 60~\gev$ \\
\midrule
\multirow{2}{*}{Low mass} & collimated dimuon & two muons with small angular separation & $\pt$ of muons $>15$ and $20~\gev$ and $\dr_{\mu\mu} < 0.5$ \\
& trimuon & three muons & $\pt > 6~\gev$ for all three muons \\
\bottomrule
\end{tabular}
}
\end{table}

The thresholds for the $\met$ and collimated-dimuon triggers changed during the course of 2016 data taking.  To account for these changes, the highest available threshold for each trigger is used and offline requirements are imposed corresponding to the thresholds listed in the table. Moreover, additional stricter requirements are imposed on the corresponding offline quantity in order to ensure that the trigger efficiency falls on the efficiency plateau.

For signal events with displaced dimuon vertices, the single-muon trigger efficiency falls off approximately linearly with $|d_0|$, from a maximum of about $70\%$ at $0~\cm$ to approximately $10\%$ at the fiducial limit of $400~\cm$, due to requirements that favor muon candidates that originate close to the IP.  The calorimeter-based $\met$ trigger is employed to recover some signal efficiency.  As muons leave little energy in the calorimeter and the $\met$ at the trigger level is computed only using the calorimeter signals, the $\met$ trigger is an effective muon trigger.  

The collimated-dimuon trigger is based on reconstruction of muon tracks with low $\pt$ thresholds.  The large rates associated with the low $\pt$ thresholds are offset by requiring two muons in the MS that are within a cone of size $\dr = 0.5$.  The efficiency of this trigger for a given signal model is strongly dependent on the magnitude of the boost of the particle decaying to the dimuon final state, as this determines the likelihood of the two muons being found within a cone of size $\dr = 0.5$.  The trimuon trigger increases the efficiency for selecting events with particles that have a relatively large branching fraction to muons, as is the case of the $\zd$ in the signal model explored in this article.

\subsection{Offline reconstruction and preselection}
\label{sec:offline}

Interaction vertices from the $pp$ collisions are reconstructed from at least two tracks with $\pt$ larger than $400~\mev$ that are consistent with originating from the beam-collision region in the $x$--$y$ plane.  Selected events are required to have at least one reconstructed interaction vertex.

Jet candidates are reconstructed from topological clusters~\cite{PERF-2014-07}, built from energy deposits in the calorimeters calibrated to the electromagnetic scale, using the anti-$k_t$ algorithm~\cite{Cacciari:2008gp} with radius parameter $R= 0.4$.  The reconstructed jets are calibrated to the hadronic energy scale by scaling their four-momenta to the particle level~\cite{PERF-2016-04}. The jets are required to have $\pt > 20~\gev$ and $|\eta |< 4.4$.  If a jet in an event fails the `Loose' jet-quality requirements of Ref.~\cite{loosejet}, the event is vetoed in order to suppress detector noise and non-collision backgrounds \cite{loosejet,DAPR-2012-01}.  To reduce the contamination due to jets originating from pileup interactions, an additional requirement on the Jet Vertex Tagger \cite{PERF-2014-03} output is made for jets with $\pt<60~\gev$ and $|\eta|<2.4$.

The muon reconstruction algorithm \cite{MUREC-1603-05598} starts by finding tracks in the MS, denoted \MuSA\ tracks, and extrapolating their trajectories towards the IP.  All \MuSA\ track parameters are expressed at the point of closest approach to the IP and their uncertainties reflect the effects of multiple Coulomb scattering in the detector material.  Although the highest track reconstruction efficiency is obtained for muons originating near the IP, appreciable efficiency is obtained for muons with transverse impact parameters as large as $200~\cm$.
In order to optimize the resolution of the track parameters, the following criteria are imposed.  The \MuSA\ tracks are required to have transverse momentum greater than $10~\gev$, $|\eta| < 2.5$, measurements in each of the three layers of both the precision and trigger chambers, an uncertainty in the $d_0$ measurement less than $20~\cm$ and to not traverse regions of the MS that are poorly aligned.  

Interactions between beam protons and beam collimators upstream of the IP are a source of high-momentum muons, denoted beam-induced-background (BIB) muons, that can enter the ATLAS detector nearly parallel to the beam axis.  Most \MuSA\ tracks generated by this process are identified and rejected with the method described in Ref.~\cite{DAPR-2012-01} and results in a negligible reduction in signal efficiency.

Track reconstruction is performed independently in the ID, and an attempt is made to match each MS track with an ID track.
The two matched tracks are then used as input to a combined fit that takes into account all of the ID and MS measurements, the energy loss in the calorimeter and multiple-scattering effects.  During the fit, additional MS measurements may be added to or removed from the track to improve the fit quality.
The ID track is required to be within the ID acceptance, $|\eta| < 2.5$, to have transverse momentum greater than $400~\mev$, to have a minimum number of hits in each ID subsystem and to have $|d_0| < 1~\cm$.  Hence, these combined-muon candidates correspond to muons produced within $\sim 1~\cm$ of the $x$--$y$ position of the IP.

To suppress background from misidentified jets as well as from hadron decays to muons inside jets, all muon candidates are required to have at least a minimum angular separation from all jets (muon--jet overlap removal) and to satisfy track-based isolation criteria.  Muon--jet overlap removal is accomplished by requiring that $\dr_{\mu-\text{jet}} > \text{min}\left( 0.4, 0.04 + 10~\gev/\pt^\mu \right )$ for all jets in the event, where $\dr_{\mu-\text{jet}}$ is the angular separation between the muon candidate and the jet in consideration.  The track-based isolation quantity $\muiso$ is defined as the ratio of the scalar sum of $\pt$ of all ID tracks matched to the primary vertex, and with $\pt > 0.5~\gev$ within a cone of size $\dr = 0.4$ around the muon candidate, to the muon $\pt$.   To remove the contribution of the ID track forming the muon candidate (if it exists), the ID track that is nearest to and within $\dr = 0.1$ of the muon candidate and has a $\pt$ within $10\%$ of the \MuSA-track $\pt$ is not used in the sum.  Muon candidates are required to have $\muiso < 0.05$.  The muon--jet overlap and isolation requirements are removed in defining control regions used to study backgrounds described in \Sect{\ref{sec:fakebkgd}}.

Muon candidates that trigger in a small set of resistive-plate chambers that can have timing jitter are rejected. This amounts to no more than $0.3\%$ of selected muon candidates, which has a negligible effect on the signal acceptance.

To distinguish between muon candidates that originate from prompt and non-prompt decays, the following classification of \MuSA\ tracks is used.  Those for which a successful ID--\MuSA\ combination has been made, defined by the requirement that the angular distance between the \MuSA\ track and nearest combined-muon track is less than $0.1$, are referred to as ``\MSID'' muon candidates and the rest are referred to as ``\MSonly'' muon candidates, as summarized in \Tab{\ref{tab:DisplacedDef}}.  The large majority of \MuSA\ tracks are \MSID, which reflects the fact that most muons are produced very close to the IP.
\begin {table}
\centering
\caption{
   \label{tab:DisplacedDef}
Definition of categories of muon candidates.  Tracks in the ID are reconstructed with maximum $|d_0|$ of $1~\cm$.}
\begin{tabular}{l  l}
\toprule
Muon candidate & Definition \\
\midrule
\MSID & successful ID--$\MuSA$ combination \\
\MSonly & standalone $\MuSA$ (no match with ID track) \\
\bottomrule
\end{tabular}

\end{table}

\subsection{Selection of dimuon vertices}
\label{sec:vertexing}

The selection criteria described below are used to define a sample of dimuon vertices (preselection) to which additional criteria are applied to form signal regions (SRs) in which data are compared to background estimates, and control regions (CRs) which are used to provide those background estimates.

Within each event, all possible pairs of muon candidates, both \MSID\ and \MSonly, are formed.  For each pair, the muon candidate with the largest $\pt$ is designated the ``leading'' muon, while the other is designated the ``subleading'' muon.  An algorithm which assumes a straight-line extrapolation of the muon trajectory from the MS inner surface towards the IP is used to determine whether or not the two muons are consistent with originating from a common vertex.  The midpoint between the points of closest approach along the trajectories of the two muon candidates is taken to be the three-dimensional location of the vertex.  This simple approach is sufficient for the purposes of this analysis, as the location of the putative dimuon vertex is only used in defining the geometrical acceptance of the analysis.  The decay length $\lvtx$ and projections onto the $x$--$y$ plane and $z$-axis, $\rvtx$ and $\zvtx$ respectively, are measured relative to the IP.  
It is convenient to sign the vertex radius $\rvtx$ according to the following definition.  If the angle between the projections in the $x$--$y$ plane of the vertex momentum vector (the dimuon momentum vector) and the ``flight direction'' (the vector connecting the IP with the displaced dimuon vertex) is less than $\pi/2$ then it is assigned a positive value, otherwise it is assigned a negative value.  When the LLP decays exclusively into a pair of muons or there is a small mass difference between the LLP and the dimuon state, the two vectors are typically closely aligned and the signed $\rvtx$ more often has a positive value.  Examples are the dark-sector model and the GGM model for cases where there is a relatively small mass difference between the $\ninoone$ and the $\Zboson$ boson.  In all cases, LLPs are distinguished by relatively large values of the magnitude of signed $\rvtx$.

Vertices are selected as follows.  To reduce combinatorial background from random track crossings, the distance of closest approach between the two straight-line extrapolations is required to be less than $20~\cm$.  As the vertex position is poorly measured for tracks that are nearly parallel to each other, vertices for which the opening angle of the muon pair is less than $0.1$ are rejected.  Vertices are required to be within the cylindrical fiducial volume $|\rvtx| <400~\cm$ and $|\zvtx| < 600~\cm$.  Background from muons with relatively low momentum in multijet events, as well as $\Upsilon$ decays to dimuons, is reduced by requiring that the dimuon invariant mass, $\mvtx$, be larger than $15~\gev$.  The ability to determine the spatial location of the vertex varies with the $\pt$ of the muons in the vertex and the opening angle between them. The average resolutions of $\rvtx$ and $\zvtx$ are in the range of $2$--$3~\cm$.
Cosmic-ray muons that pass through the detector in time with a $pp$ collision are sometimes reconstructed as two separate $\MuSA$ tracks that have an opening angle of $\pi$:  $\Delta \phi = \pi$ and $\Sigma \eta = 0$, where $\Delta \phi$ is the difference in $\phi$ between the two $\MuSA$ tracks and $\Sigma \eta$ is the sum of their $\eta$ values.  Vertices formed by such $\MuSA$ tracks are effectively eliminated by requiring $\sqrt{(\Sigma \eta)^{2} + (\pi - \Delta \phi)^{2}} > 0.1$.

Backgrounds that contribute to the preselection sample include SM proton-proton collision processes as well as events with muons that are not associated with the $pp$ collision (e.g.\ cosmic-ray muons).  The dominant contributions to the former are low-mass Drell--Yan and $\Zboson+\text{jets}$ processes, collectively referred to simply as DY.  At small values of $\mvtx$, dimuon vertices from multijet processes are also substantial.  Dimuon vertices reconstructed in $\ttbar$ and single-top events make small contributions, while $W+\text{jets}$ and diboson processes are found to be negligibly small backgrounds.

\begin{figure}[htbp!]
\centering
\subfloat[]{\includegraphics[width = 0.49\textwidth]{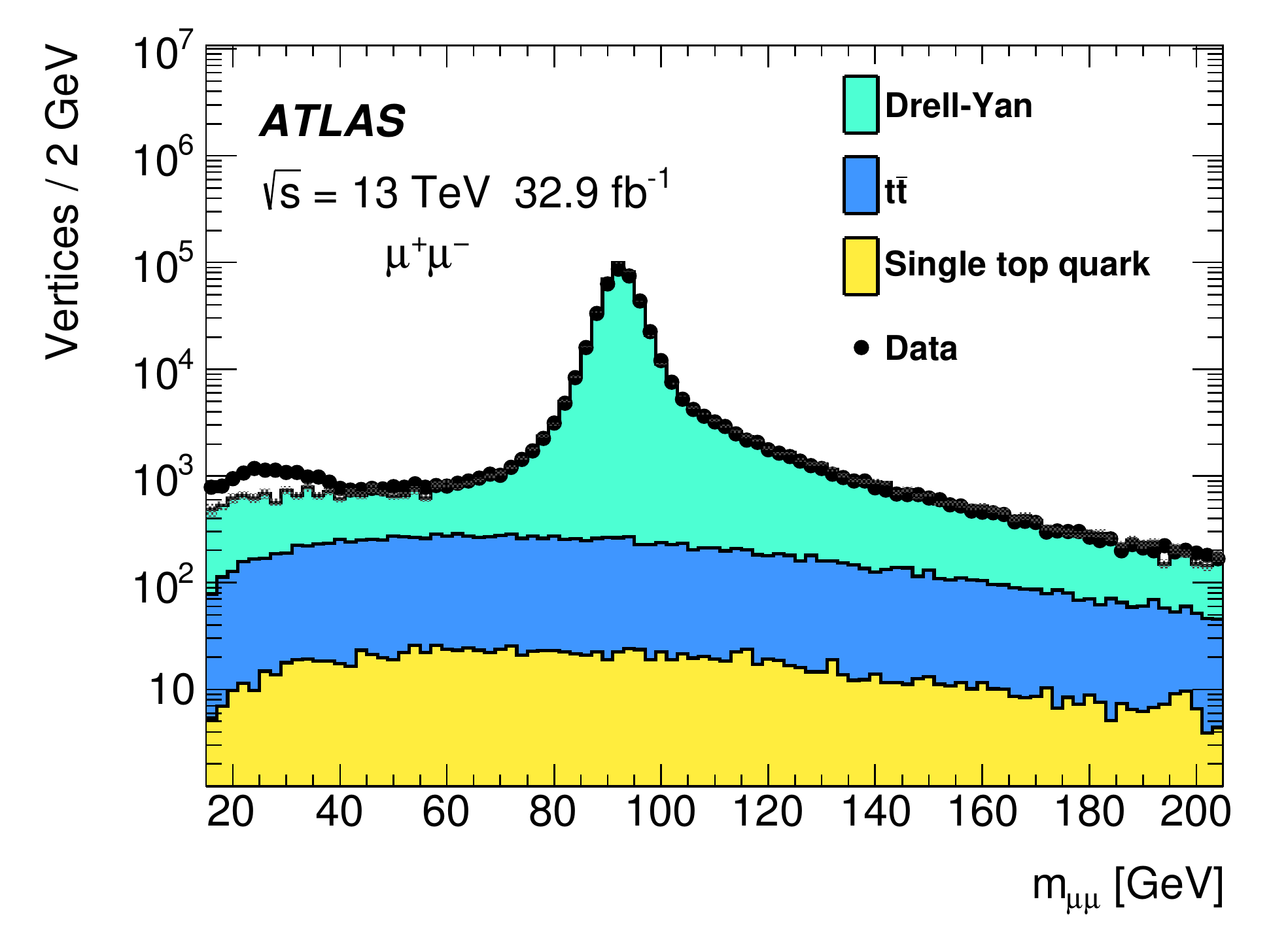}}
\subfloat[]{\includegraphics[width = 0.49\textwidth]{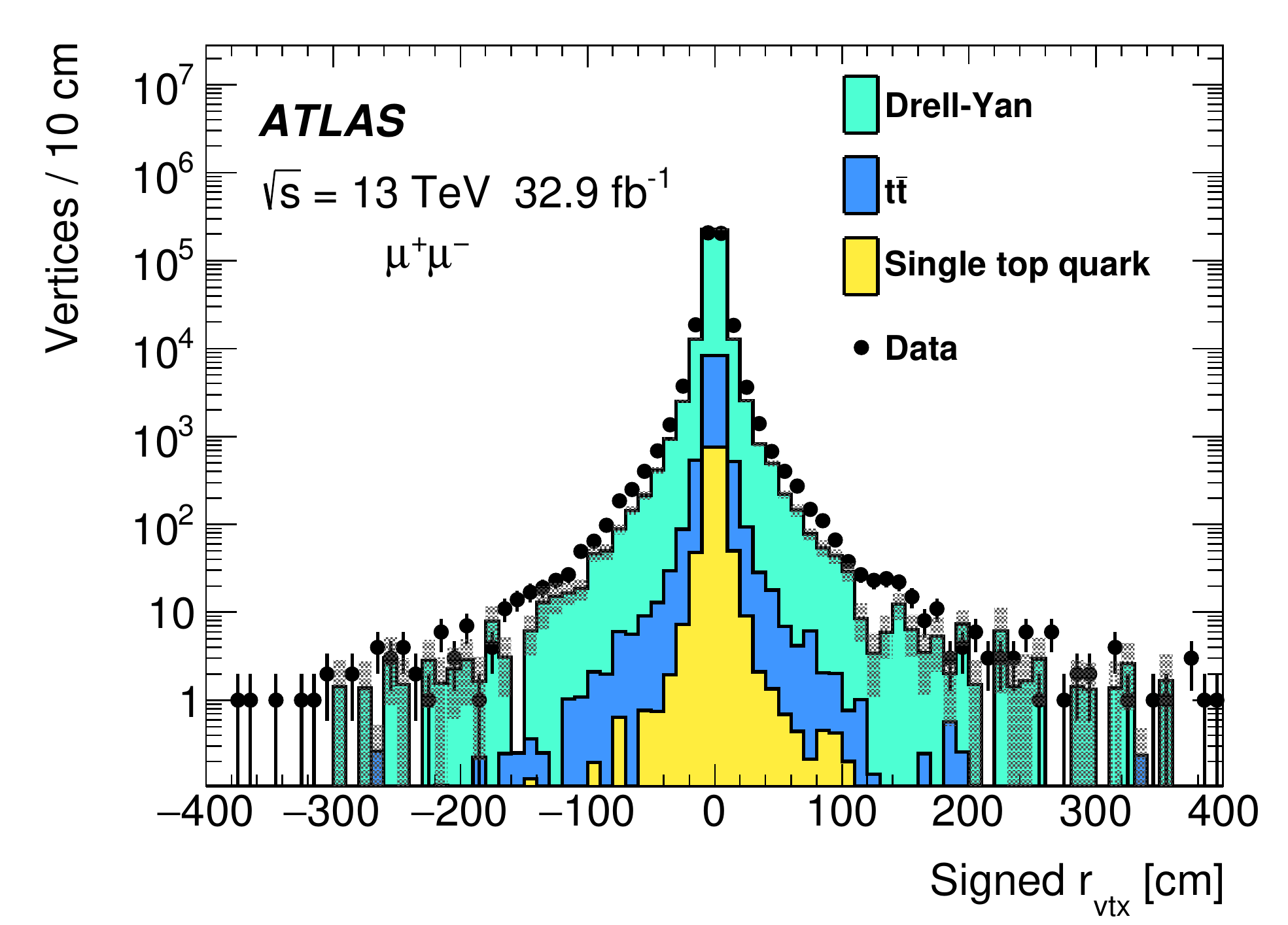}} \\
\subfloat[]{\includegraphics[width = 0.49\textwidth]{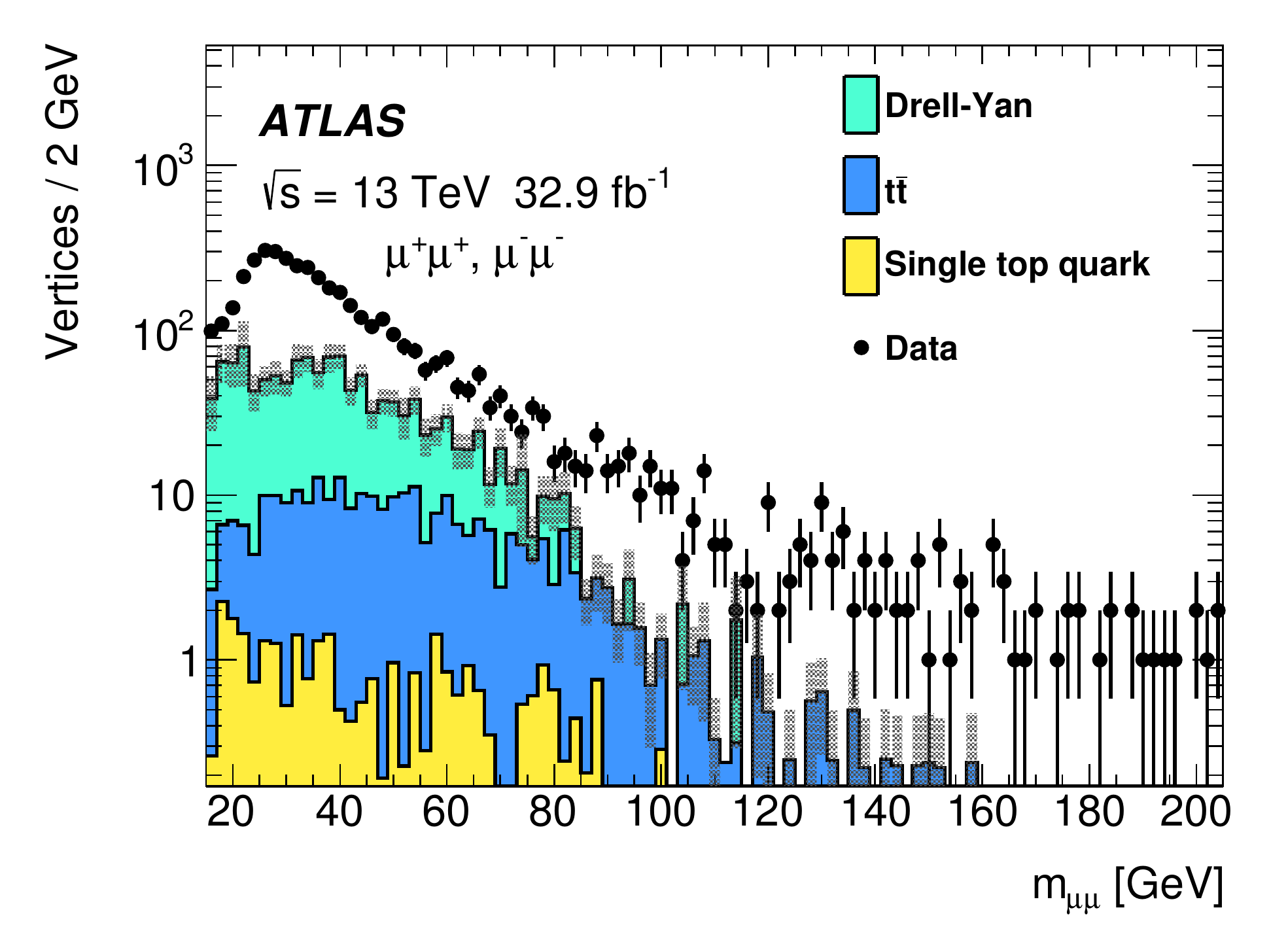}}
\subfloat[]{\includegraphics[width = 0.49\textwidth]{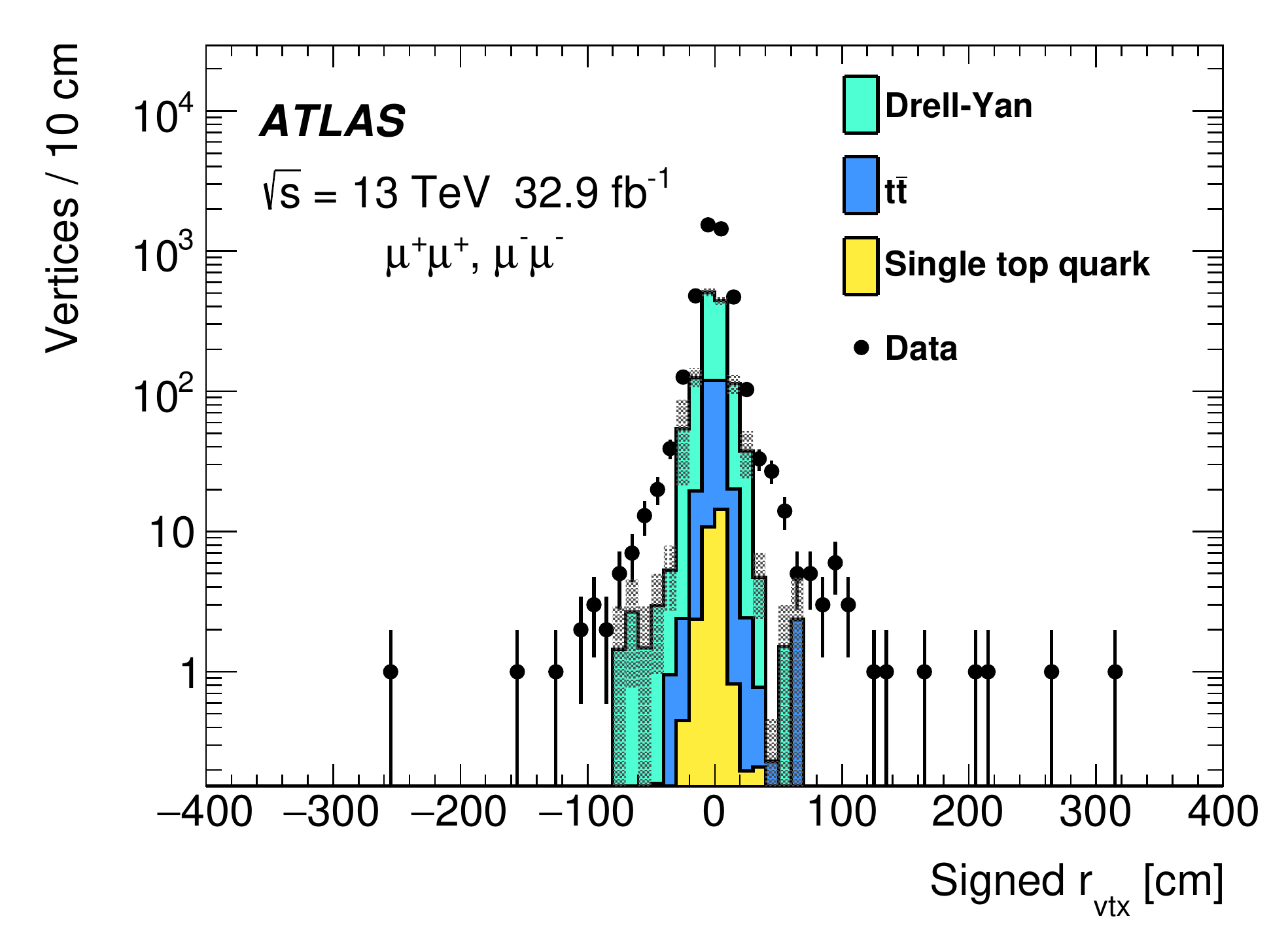}}
\caption{
	\label{fig:smCRPlots}
Distributions of: (a) dimuon invariant mass $\mvtx$ and (b) signed vertex radius $\rvtx$ for opposite-charge dimuon vertices satisfying the preselection requirements described in the text; 
(c) $\mvtx$ and (d) signed $\rvtx$ for same-charge dimuon vertices satisfying the preselection requirements described in the text.
The stacked histograms represent the expected contributions from various SM background processes and are derived from MC simulations scaled to an integrated luminosity of $\lumi$.  Multijet processes are not included in the background due to the limited number of simulated events.  The contributions from these processes are most substantial at small values of $\mvtx$.  The shaded bands represent the statistical uncertainties in the simulated background.  The observed distributions for data are given by the points with error bars.}
\end{figure}

Distributions of $\mvtx$ and signed $\rvtx$ for opposite-charge and same-charge (SS) dimuon vertices satisfying the preselection criteria are shown in \Fig{\ref{fig:smCRPlots}}.  Also shown are the expected contributions from the background processes discussed above.  Due to the limited number of simulated multijet events, this source of background is not included in the MC distributions.  Its relative contribution is expected to be dominant for SS pairs and most pronounced for OS ones at small values of $\mvtx$, as determined from studies of events where the muon--jet overlap and muon isolation requirements are inverted, and this is the dominant source of difference between the data and MC distributions in those regions of \Fig{\ref{fig:smCRPlots}}.
The fraction of events in the data with multiple dimuon vertices passing the preselection criteria is $0.065\%$. 

The preselected dimuon vertices are divided into two regions to be used in searches for low- and high-mass signal models, which are summarized in \Tab{\ref{tab:Massdefs}}.  To further suppress the DY background in the high-mass region, where $\Zboson + \text{jets}$ production dominates, and improve the search sensitivity, the transverse boost of the dimuon pair, defined as the ratio of the transverse momentum of the dimuon system to its invariant mass, is required to be larger than two.  This reduces the DY background by approximately a factor of $20$, with a small reduction in the signal efficiencies, where the decay of a heavy BSM particle produces the dimuon state (a $\Zboson$ boson in the GGM model) with a relatively large boost.   

\begin {table}
\centering
\caption{
   \label{tab:Massdefs}
Selection criteria for low- and high-mass regions, in addition to the preselection requirements described in the text.  The definitions of the low- and high-mass signal regions are also given.}\begin{tabular}{l  l  l }
\toprule
Selection & Low mass & High mass \\
\midrule
$\pt^\mu~[\gev]$ & $>10$ & $>20$ \\
$\mvtx~[\gev]$ & $15$--$60$ & $>60$ \\
Dimuon transverse boost & -- & $> 2$ \\
\midrule
          & $\srzd$ & $\srggm$ \\
\midrule
Muon candidates & both \MSonly & both \MSonly \\
Muon candidate charge & opposite charge & opposite charge \\
\bottomrule
\end{tabular}

\end{table}

The next sections describe the SR and CR selection criteria based on the designation of muon candidates as \MSID\ or \MSonly.

\subsection{Signal regions and signal efficiency}
\label{sec:sr}

Signal is characterized by vertices where both muon candidates are \MSonly.  This requirement unavoidably leads to a reduction in efficiency for decays close to the IP.  Displaced-vertex analyses that make use of ID tracks \cite{SUSY-2014-02} effectively recover such signal events.  Two orthogonal signal regions are used to increase the sensitivity to low- and high-mass signal models, $\srzd$ and $\srggm$, respectively.  The two regions are summarized in \Tab{\ref{tab:Massdefs}}.  For both SRs, the muons are required to have opposite charge.

\begin{figure}[!htbp]
\centering
\subfloat[]{\includegraphics[width = 0.49 \textwidth]{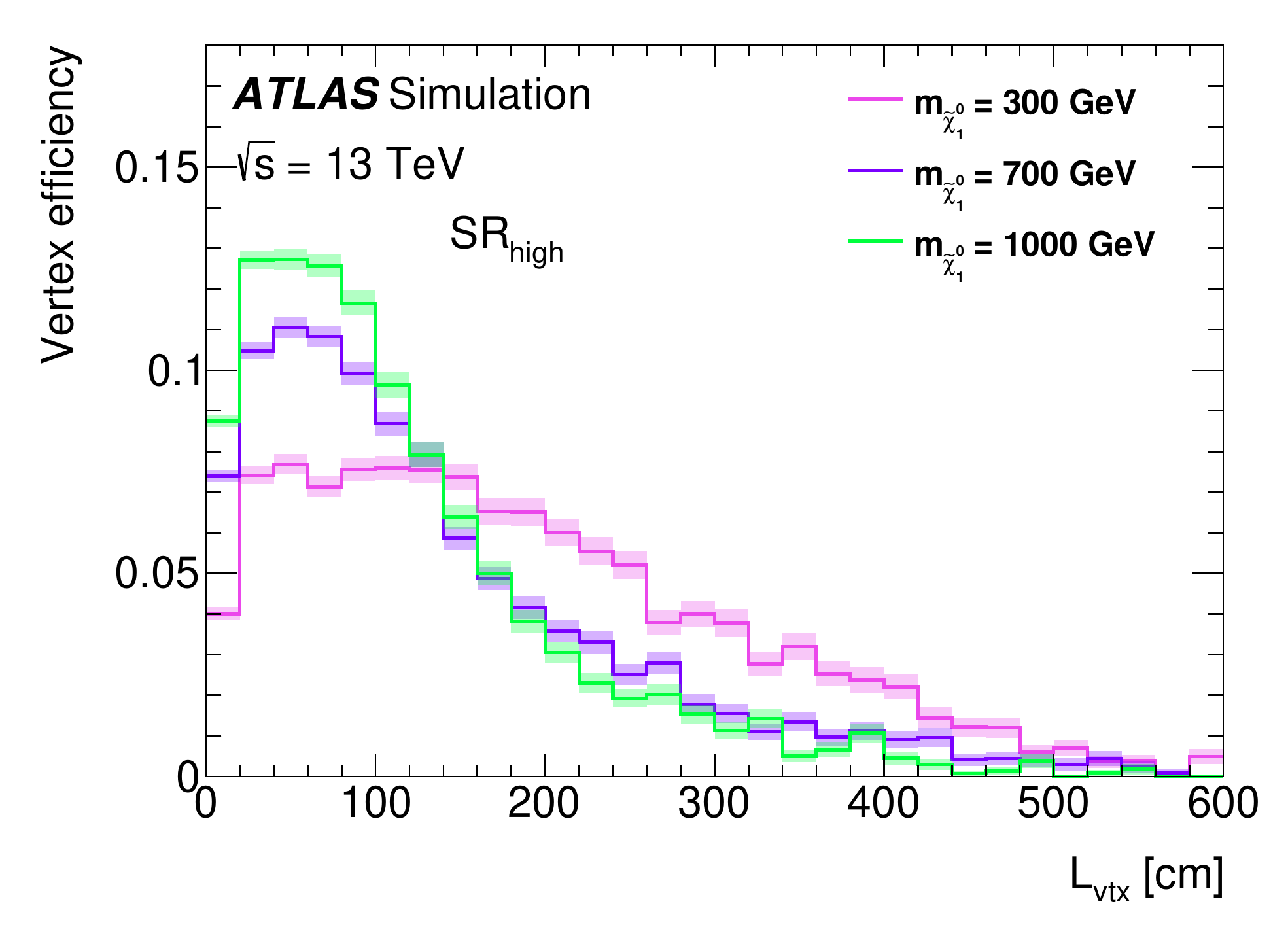}}
\subfloat[]{\includegraphics[width = 0.49 \textwidth]{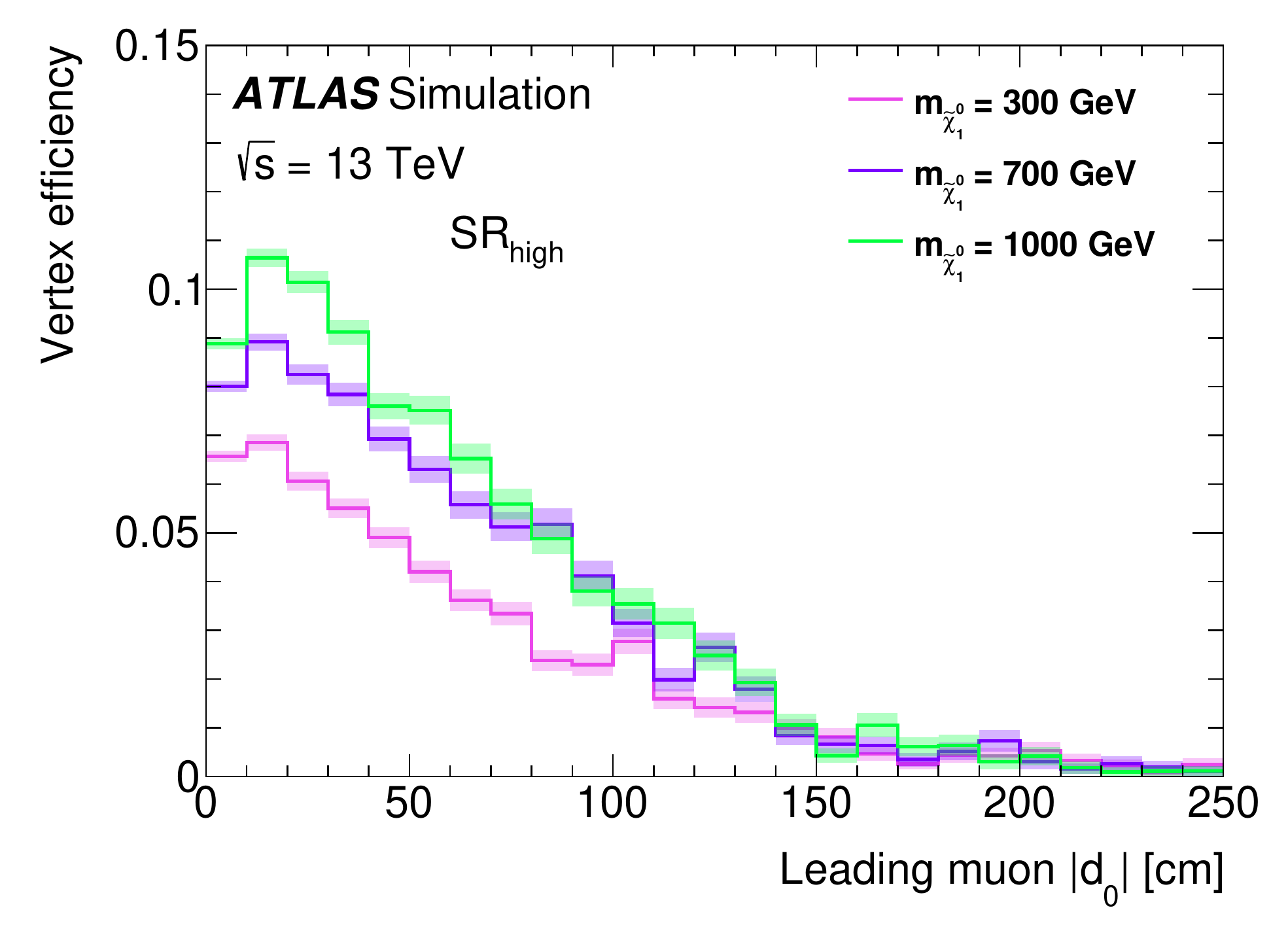}} \\
\subfloat[]{\includegraphics[width = 0.49 \textwidth]{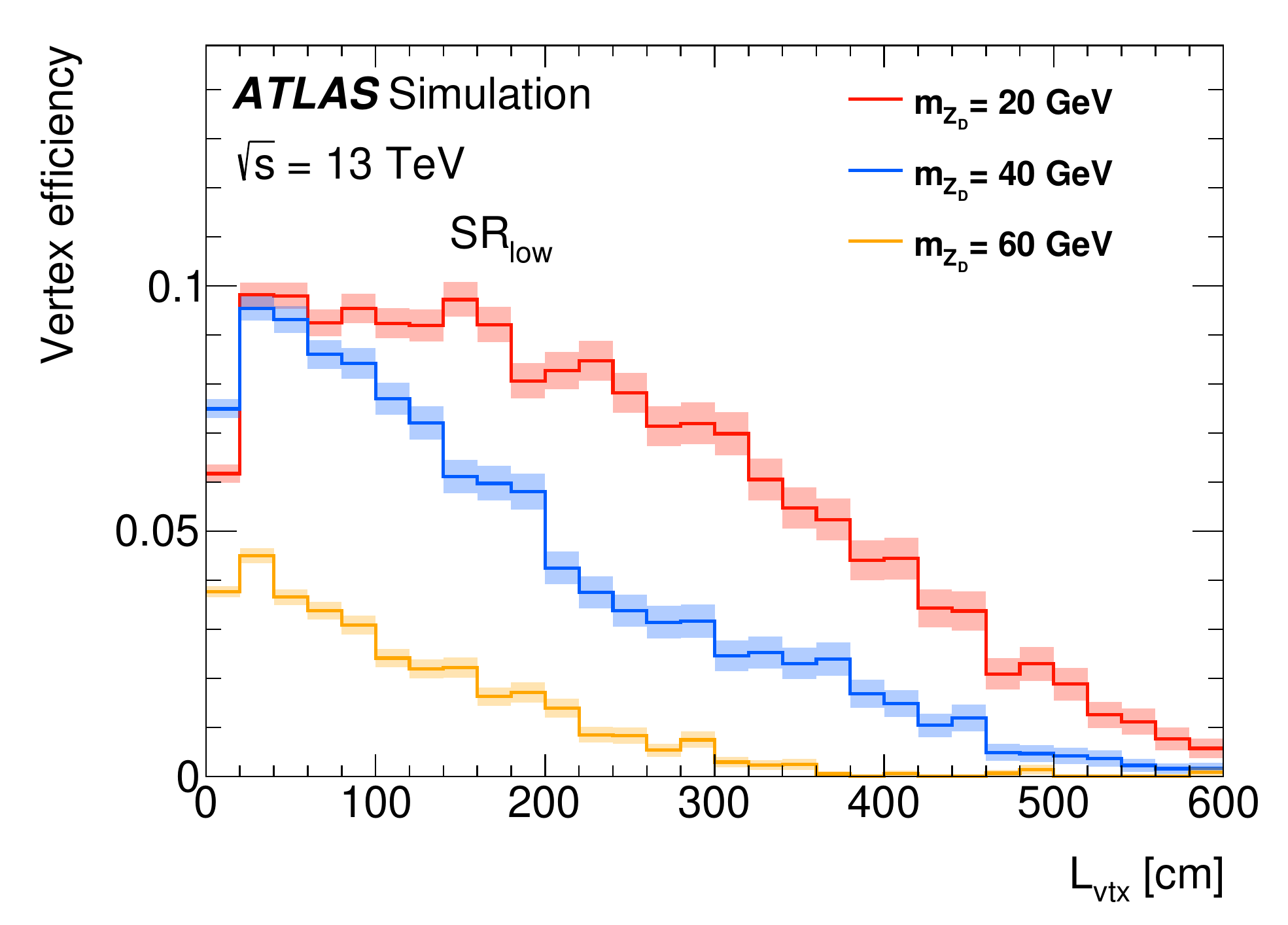}}
\subfloat[]{\includegraphics[width = 0.49 \textwidth]{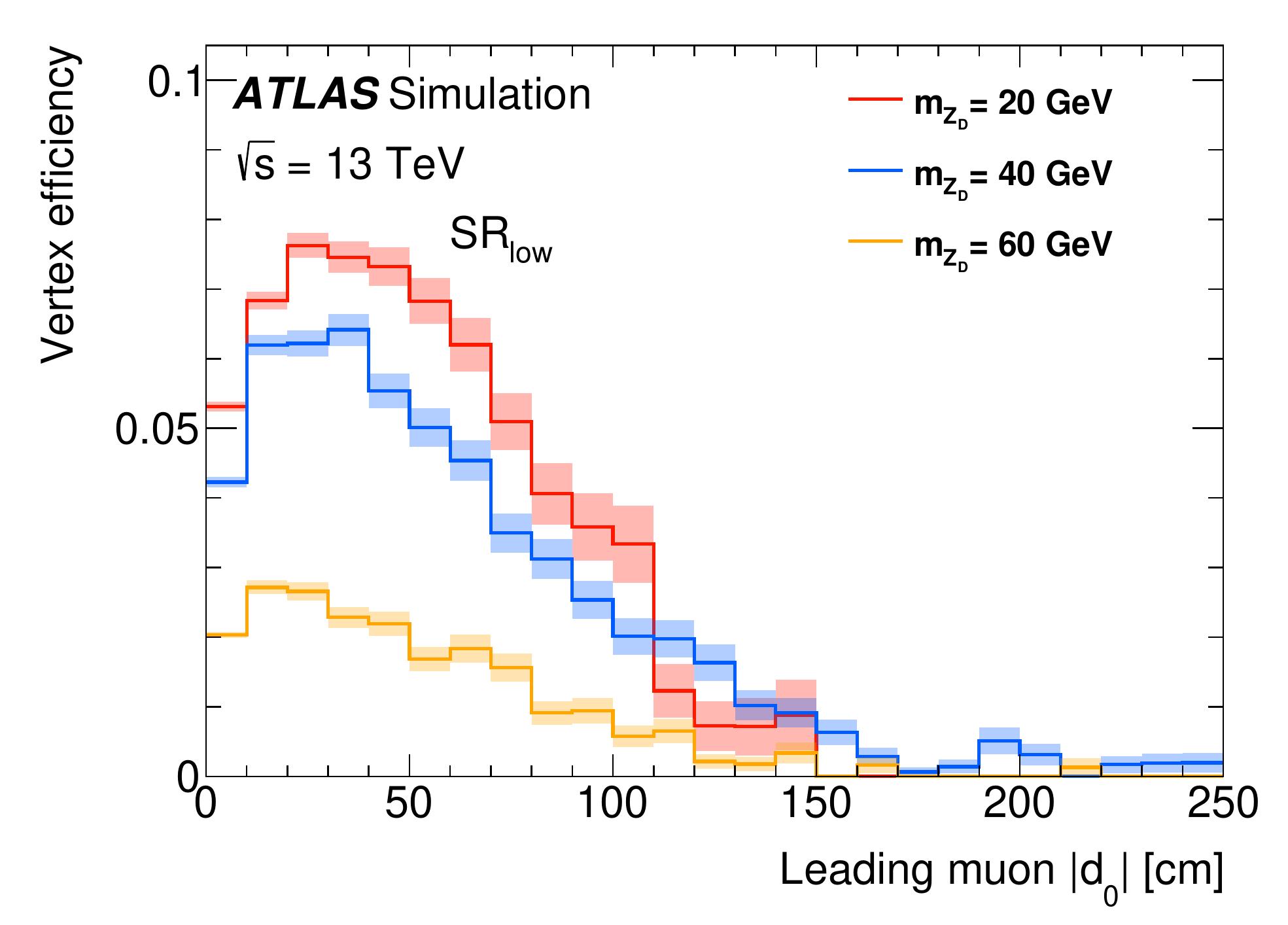}}

\caption{
	\label{fig:effvsrz}
The efficiency for selecting a displaced dimuon vertex that satisfies the requirements of $\srggm$ and $\srzd$ as function of (a) and (c) generated decay length $\lvtx$, and (b) and (d) generated transverse impact parameter $d_{0}$ of the leading muon. These efficiencies are calculated relative to all generated signal vertices and include geometrical acceptance and reconstruction effects.  The distributions in (a) and (b) are derived from signal events with a long-lived neutralino, $\ninoone$, decaying to a $\Zboson$ boson (with $\zmumu$) and a gravitino.  The distributions in (c) and (d) are derived from signal events with a long-lived dark photon, $\zd$, that decays to $\mu^+\mu^-$.  The shaded bands represent the statistical uncertainties in the efficiencies.
}
\end{figure}

The product of acceptance and reconstruction efficiency determined from simulated signal events is shown in \Fig{\ref{fig:effvsrz}} as a function of generated $\lvtx$ and leading muon $d_{0}$, for the GGM model and for the dark-sector model.  The lower efficiency observed for small $\lvtx$ or small $|d_0|$ (more apparent in the $\zd$ models) is due to the veto on \MSID\ muons, while the loss at large values reflects the lower \MuSA\  reconstruction efficiency for tracks with trajectories that do not extrapolate back to a region close to the IP.  
The value of $\lvtx$ where maximum efficiency is achieved is different for each choice of $\zd$ mass due to the large differences in boost.  

\begin{figure}[!htbp]
\centering
\subfloat[]{\includegraphics[width = 0.49 \textwidth]{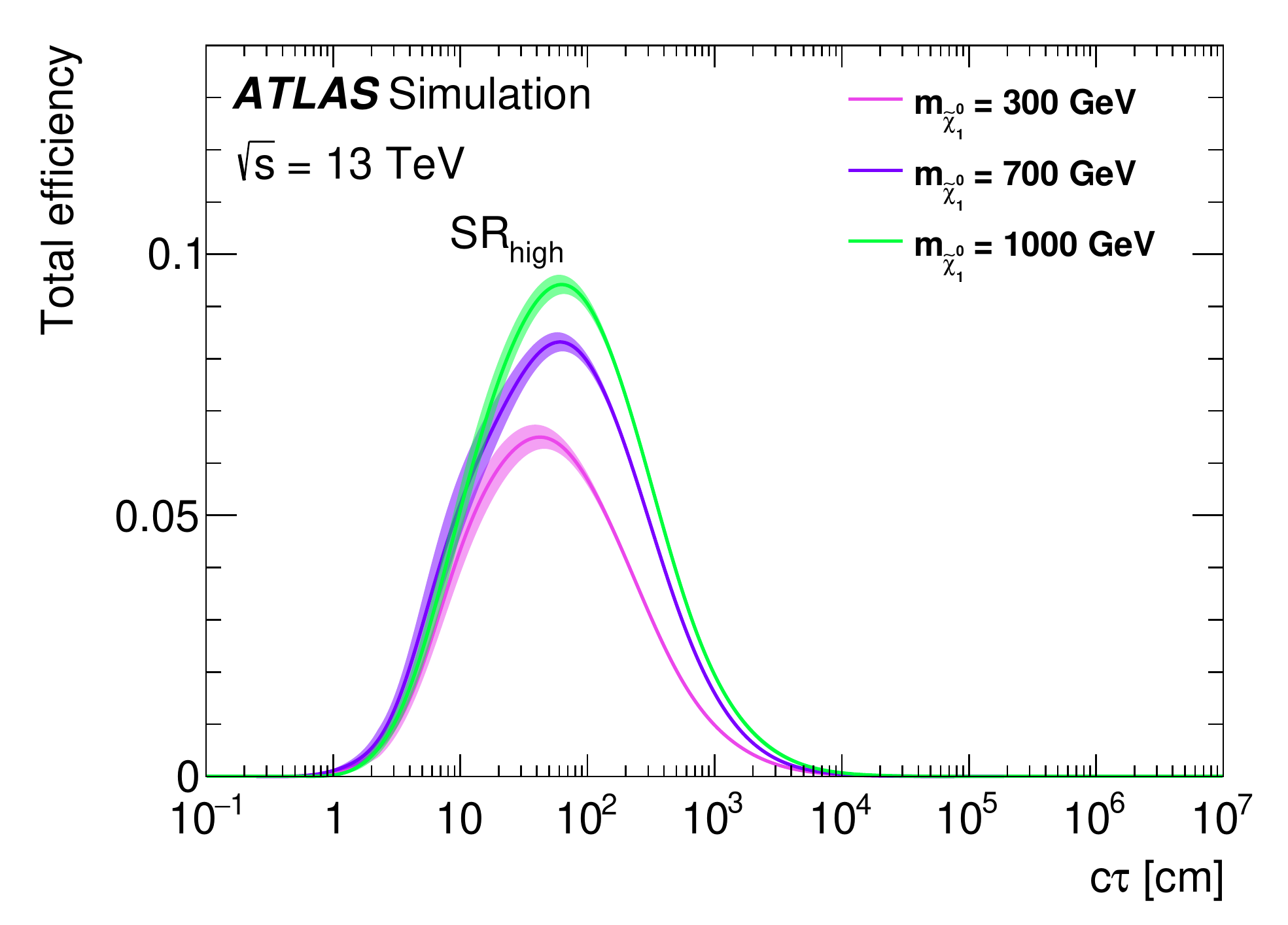}} 
\subfloat[]{\includegraphics[width = 0.49 \textwidth]{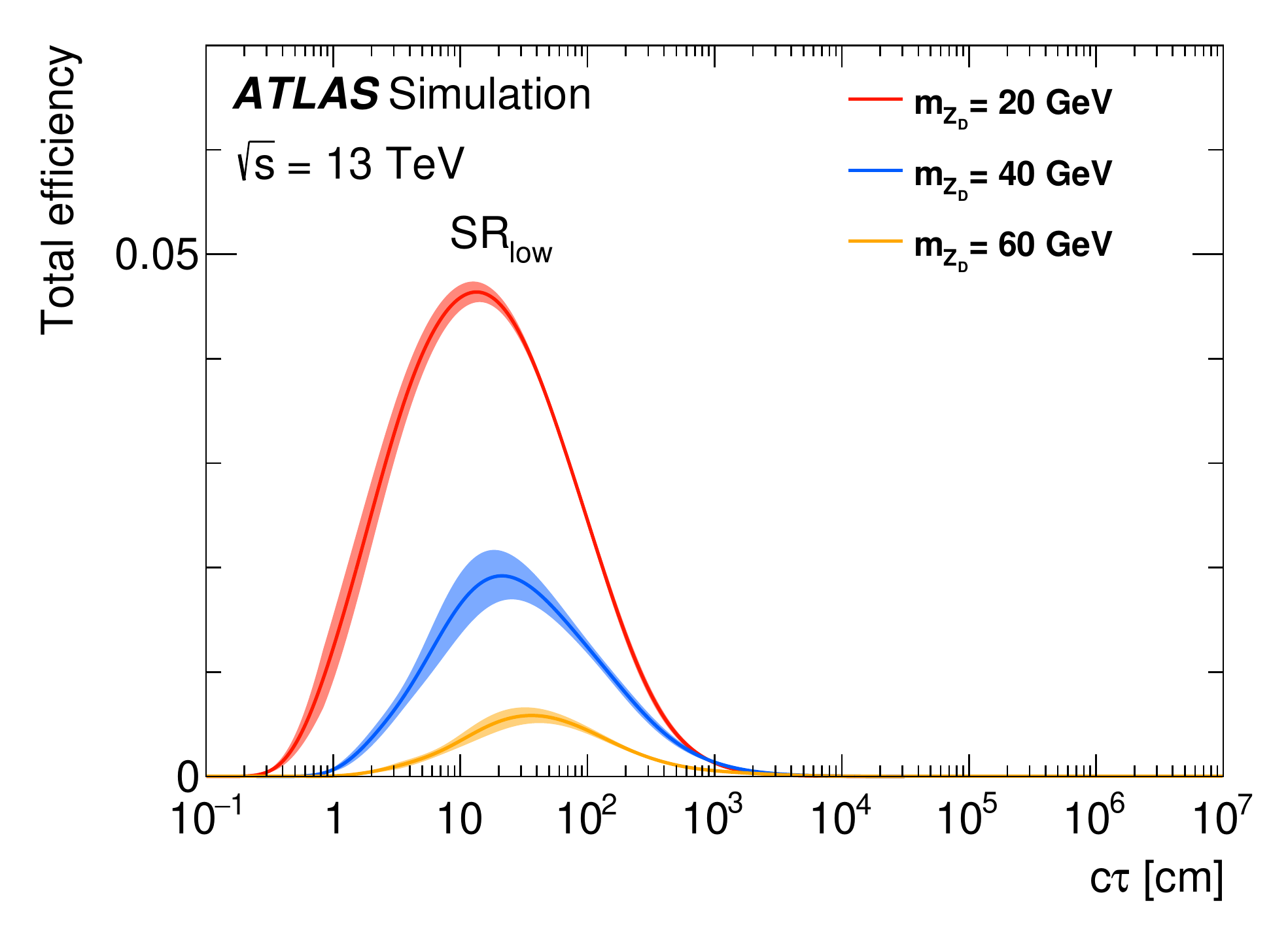}} 
\caption{
    \label {fig:efflifetimes}
Overall event-level efficiencies after the signal-region selections (combining trigger and offline selection), as a function of the lifetime of the long-lived BSM particle, for (a) the GGM model and (b) the dark-sector model. The shaded bands represent the statistical uncertainties only.}
\end{figure}

The total event-level efficiencies, including trigger and offline selection criteria, as functions of the lifetime of the LLP, are shown in \Fig{\ref{fig:efflifetimes}} and maximum values are in the $c\tau$ region $20$--$50~\cm$.  The reweighted samples, as described in \Sect{\ref{sec:mcsamples}}, are used to estimate the efficiencies for values of the lifetime which were not used in generating the simulated samples.  This event-level efficiency is defined as the fraction of generated events that are selected and have at least one dimuon DV.

\begin{figure}[!htp]
\centering
\subfloat[]{\includegraphics[width = 0.49 \textwidth]{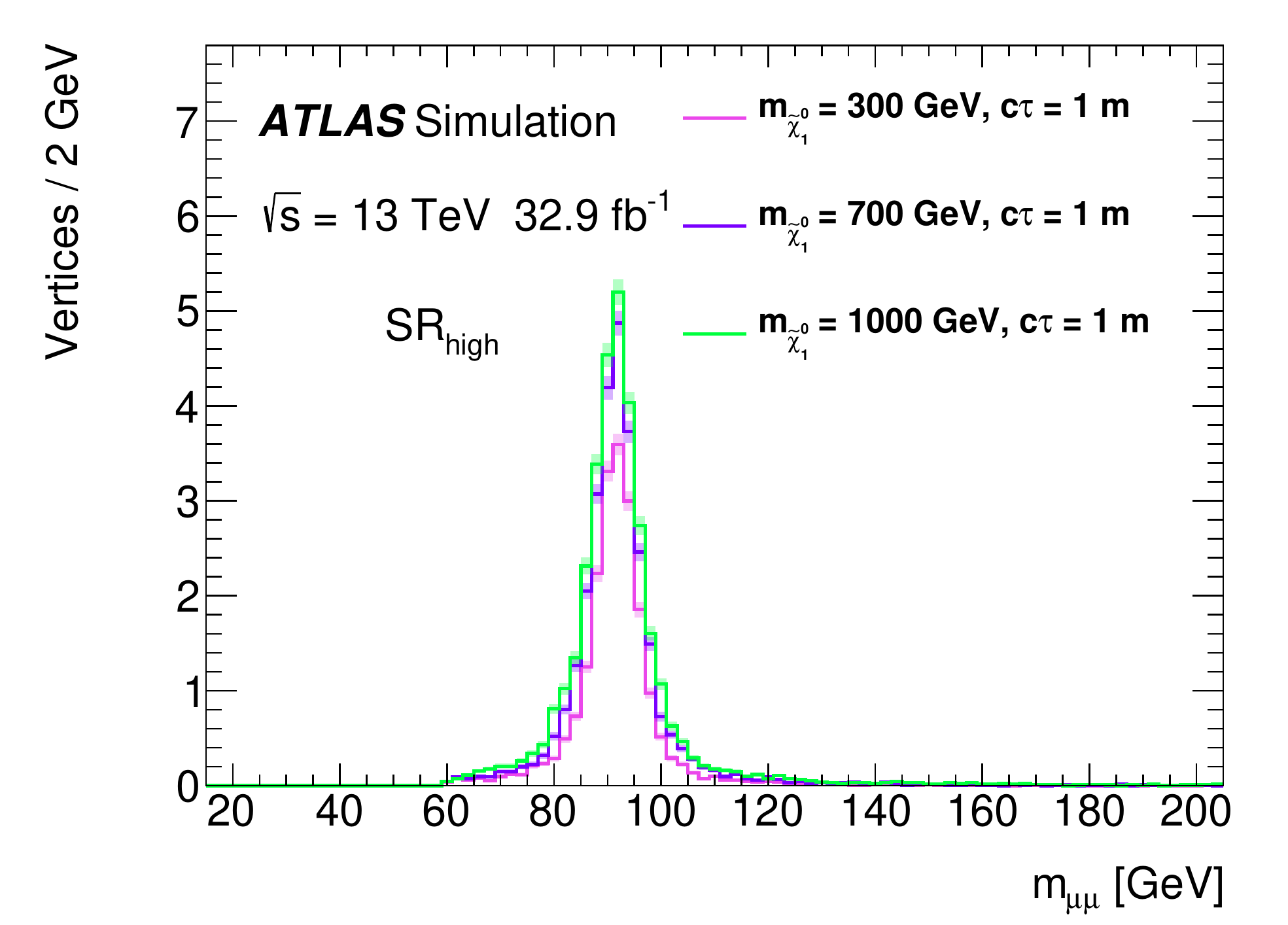}}  
\subfloat[]{\includegraphics[width = 0.49 \textwidth]{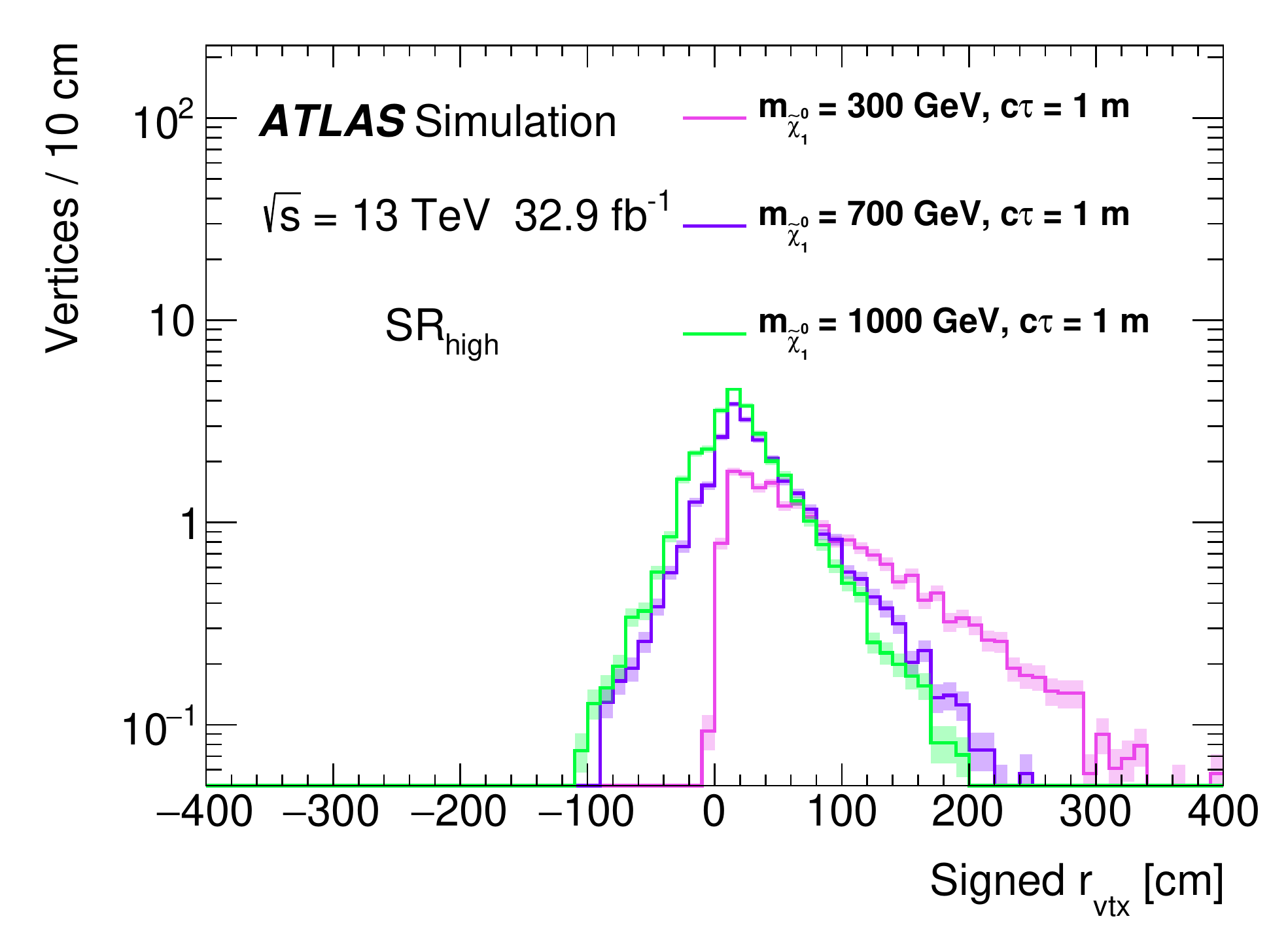}} \\
\subfloat[]{\includegraphics[width = 0.49 \textwidth]{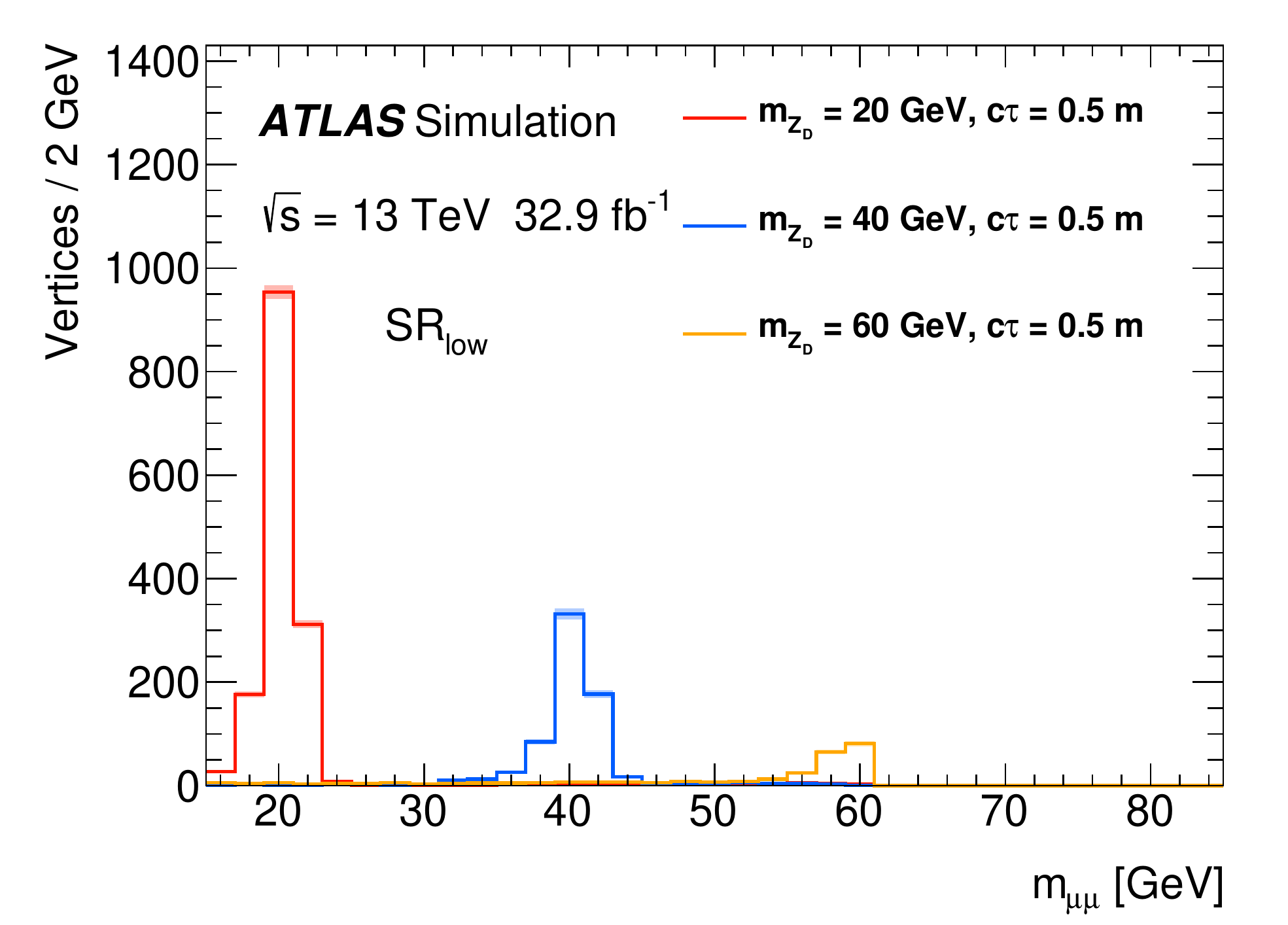}}  
\subfloat[]{\includegraphics[width = 0.49 \textwidth]{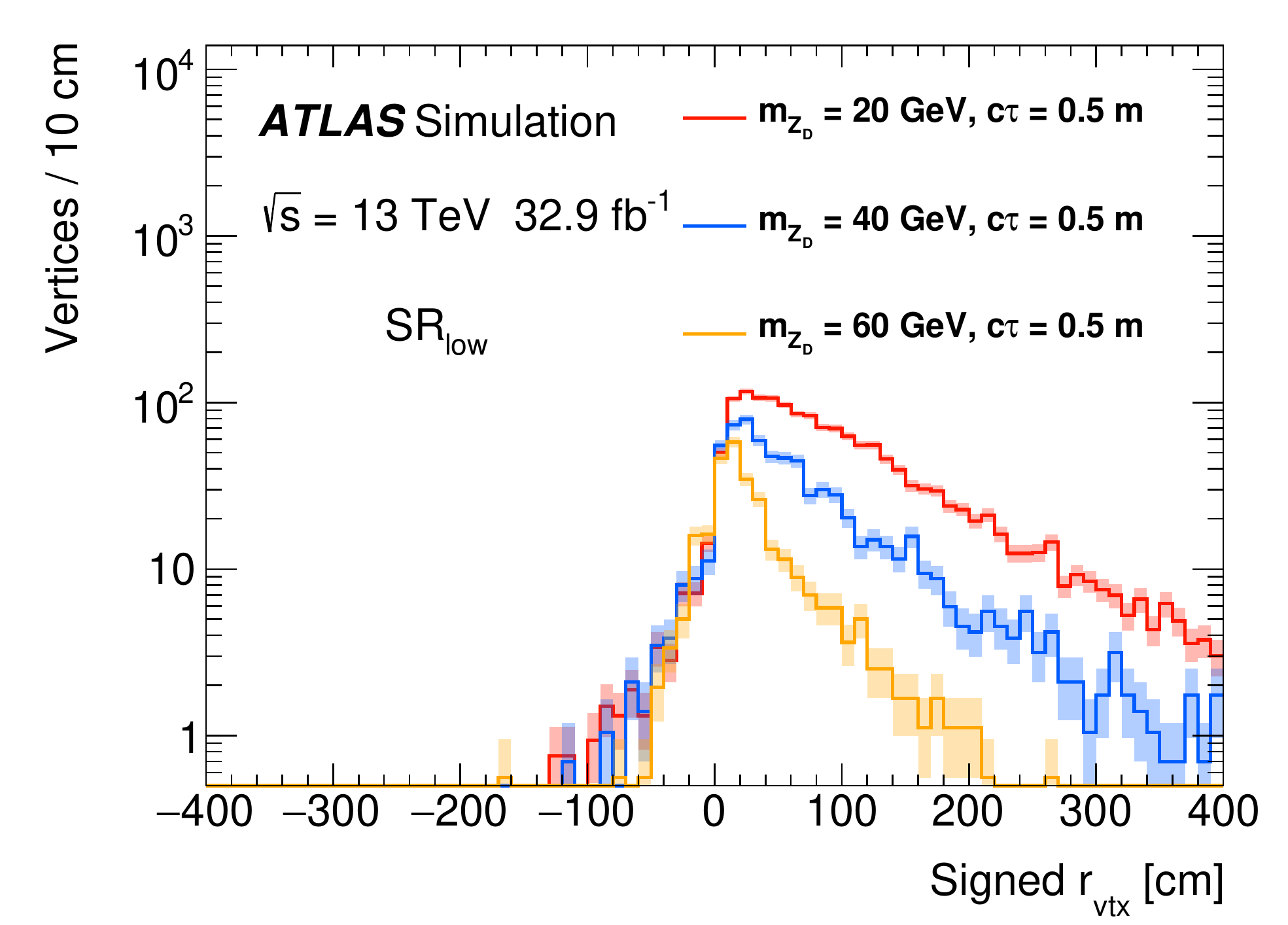}}
\caption{
    \label {fig:SRall}
    Distributions derived from MC simulations of:  (a) dimuon invariant mass $\mvtx$ and (b) vertex radius $\rvtx$ for signal vertices in $\srggm$ with a long-lived neutralino, $\ninoone$ ($\mchi=300,~700,~\text{and}~1000~\gev$ and $\ctchi = 100~\cm$) decaying to a $\Zboson$ boson (with $\zmumu$) and a gravitino; (c) $\mvtx$ and (d) $\rvtx$ for signal vertices in $\srzd$ with a long-lived dark photon, $\zd$ ($\mzd=20,~40,~\text{and}~60~\gev$; and $\ctzd = 50~\cm$), that decays to $\mu^+\mu^-$. The shaded bands represent the statistical uncertainties.  The distributions are normalized to the expected yields in the signal regions for:  $m_{\gluino}=1100~\gev$, $\sigma\left(pp\to \gluino\gluino \right) = 0.1635~\pb$, $B\left(\ninoone \to \Zboson \gravino \right) = 1.0$ and $B\left(\Zboson \to \mu^+\mu^- \right) = 0.03366$; and $m_{\Hboson}=125~\gev$, $m_{\Hd}=300~\gev$, $\sigma\left(pp\to \Hboson \right) = 44.1~\pb$, $B\left( \Hboson \to \zd\zd \right) = 100\%$, and the value of $B\left(\zd\to \mu^+\mu^-\right)$ varying between $0.1475$ and $0.1066$ for the range $\mzd = 20$--$60~\gev$.
}
\end{figure}

Distributions of $\mvtx$ and signed $\rvtx$ for signal vertices in simulated events, for both $\srggm$ and $\srzd$, are displayed in \Fig{\ref{fig:SRall}}. The vertex properties are computed using the parameters of the reconstructed MS tracks and the distributions are normalized to the expected yields in the signal regions.

\subsection{Control regions and background estimation}
\label{sec:cr}

Dimuon vertices are categorized as described in \Tab{\ref{tab:ABCDdefs}}.
The observed yields of same-charge dimuon vertices in all four regions A, B, C and D are used to estimate the background yields in the SRs due to muons produced more than about a centimeter from the IP, referred to as non-prompt muons.  The observed yields in the opposite-charge B, C and D CRs are used to predict the background yield from SM processes that produce prompt muons (those produced within about a centimeter of the IP) in the SRs (opposite-charge dimuon vertices in region A).  Muons from decays of hadrons containing $b$ and $c$ quarks are, within the context of this analysis, considered to be prompt muons.
\begin {table}[!h]
\centering
\caption{
   \label{tab:ABCDdefs}
Description of four regions used in estimating background yields.  The ordering of the muon candidates in the description of the dimuon vertex is leading muon first, then subleading muon.}\begin{tabular}{l  l }
\toprule
Region name & Muon candidates in vertex \\
\midrule
A & \MSonly --\MSonly \\
B & \MSonly --\MSID \\
C & \MSID --\MSonly \\
D & \MSID --\MSID \\
\bottomrule
\end{tabular}
\end{table}

\subsection{Non-prompt muon vertices}
\label{sec:fakebkgd}

Non-prompt muons are those for which no matching ID track is expected.  Examples of such sources of background non-prompt muons are cosmic-ray muons, BIB muons, fake \MuSA\ tracks generated from random hit combinations, and those arising from pion or kaon decay.  

Cosmic-ray and BIB muons will usually not produce ID tracks, as they rarely pass close enough to the IP to produce an ID track that satisfies the track reconstruction criteria, in particular the $|d_0| < 1~\cm$ requirement.
As a result, they produce mostly \MSonly\ muon candidates.  As described in \Sect{\ref{sec:offline}}, dimuon vertices reconstructed from a single cosmic-ray or BIB muon that generates two $\MuSA$ tracks are effectively eliminated in the preselection by taking advantage of the fact that the angle between the two tracks will be nearly $180^\circ$.  On the other hand, vertices formed when a single $\MuSA$ track from a cosmic-ray or BIB muon is paired with an unrelated muon candidate produced from the $pp$ collision will satisfy these selection criteria and more readily contribute to the background.

Pions and kaons have relatively large lifetimes and feature large branching fractions to final states with one muon.  Such decays often result in either no ID track being reconstructed, due to the requirement of at least a minimum number of ID hits, or the ID track of the pion/kaon failing to be matched to the muon \MuSA\ track.  In both of these two cases, a \MSonly\ muon candidate will be produced.

Vertices that contain one or more non-prompt muons are referred to as ``non-prompt vertices."  If the vertex contains a cosmic-ray or BIB muon paired with an unrelated muon candidate, the charges of the two $\MuSA$ tracks will be largely uncorrelated. However, some charge correlation is expected in vertices containing muons from pion/kaon decay, because the pion/kaon is produced from the same $pp$ collision that produces the other muon in the vertex.  For the latter, the charge correlation is studied by measuring the ratio of OS to SS dimuon vertices, $\Rqqcd$, in the data.  As muons from pion/kaon decay are not expected to be isolated from jets, the quantity $\Rqqcd$ is measured in a sample of dimuon vertices where the selection criteria are those of the preselection, except that the jet--muon overlap and isolation requirements are removed for both muons. 
The value ranges from $1.39\pm 0.09$ to $1.55\pm 0.03$, depending on the region (A, B, C or D).  As no statistical difference in $\Rqqcd$ is observed for the low- and high-mass regions, a single value is used for both regions.  For dimuon vertices composed of BIB or cosmic-ray muons, the muon charges are assumed to be entirely uncorrelated:  $\Rqcos = 1.0$.

Since the relative composition of the non-prompt dimuon background is unknown, the average of $\Rqqcd$ and $\Rqcos$ is assumed as the nominal value, $\Rq$, with an uncertainty that is half the difference between the two values; this is shown in Table~\ref{tab:SRfakeYields} ($\Rq = 1.24 \pm 0.24$ for the SRs).

The numbers of OS non-prompt vertices in the regions A, B, C and D  are predicted using the number of SS vertices in each region and the appropriate $\Rq$, as described above:  $\nfake_i = R_{q,i} \nss_i$, where $\nss_i$ is the number of SS vertices observed in region $i$ ($i = $A, B, C or D) and $R_{q,i}$ is the charge ratio for region $i$.  

The predicted yields of non-prompt dimuon vertices for both SRs are given in \Tab{\ref{tab:SRfakeYields}}, where the uncertainty in $\Rq$ is treated as a systematic uncertainty added in quadrature to the statistical uncertainties. 
\begin {table}[htb!]
\centering
\caption{
   \label{tab:SRfakeYields}
Observed number of SS vertices $\nss$, charge ratio $R_{q,\text{SR}}$ and predicted yields of non-prompt dimuon vertices $\nfake_\text{SR}$ in each signal region. The uncertainty in $\nfake_\text{SR}$ combines in quadrature the statistical uncertainty, derived from the observed number of SS vertices, and the uncertainty in the charge ratio.  The statistical uncertainty on $\nss$ for the case where the central value is measured to be zero is taken to be the $68\%$ one-sided Poisson confidence-level interval: $+1.14$.}
\begin{tabular}{ l  c c c }
\toprule
Region & $\nss$ & $R_{q,\text{SR}}$ & $\nfake_\text{SR}$ \\[+2pt]
\midrule
\srzd & $11$ & $1.24 \pm 0.24$ & $13.6 \pm 4.9$ \\

\srggm & $0$ & $1.24 \pm 0.24$ & $0.0^{+1.4}_{-0.0}$ \\
\bottomrule
\end{tabular}
\end{table}

\subsection{Prompt muon vertices}
\label{sec:smbkgd}
 
The number of dimuon vertices in each of the SRs arising from prompt muon processes is estimated from the observed yields in the OS low-mass and high-mass B, C and D control regions.  
Sources of such background in the SRs include SM processes that produce prompt muons that are reconstructed as \MSonly\ due to detector or reconstruction effects, such as ID inefficiencies, or poorly reconstructed combined muons, collectively described as failed combined muon reconstruction.

To avoid double-counting of dimuons from non-prompt processes, the estimated number of non-prompt OS vertices in each region is subtracted:
\begin{equation*}
 \nprompt_i = \nos_i - \nfake_i, ~ ~ ~ ~ ~ i = \text{B},\; \text{C},\; \text{D}\, ,
\end{equation*}
where $\nos_i$ is the number of OS vertices in region $i$, $\nfake_i$ is the number of OS non-prompt vertices in region $i$ (described in \Sect{\ref{sec:fakebkgd}}) and $\nprompt_i$ is the estimated number of opposite-charge vertices from prompt processes in region $i$.
The quantity $\nprompt_\text{B}$ ($\nprompt_\text{C}$) is the estimated number of OS vertices from prompt processes with leading (subleading) muons that fail the combined reconstruction and are identified as \MSonly, while the other muon candidate is identified as being \MSID.  The quantity $\nprompt_\text{D}$ is the estimated number of OS vertices from prompt processes with muon candidates that pass the combined reconstruction and are both identified as being \MSID.  With these definitions, the leading and subleading ``transfer factors'' are defined as follows:
\begin{eqnarray*}
  f_\text{L} = {\nprompt_\text{B}}/{\nprompt_\text{D}}\, , \\
  f_\text{S} = {\nprompt_\text{C}}/{\nprompt_\text{D}}\, .
\end{eqnarray*}

The leading transfer factor multiplied by $\nprompt_\text{C}$, or, alternatively, the subleading transfer factor multiplied by $\nprompt_\text{B}$, thus gives for prompt muon processes the predicted number of OS vertices in region A, the SRs in this case:
\begin{equation*}
\nprompt_\text{A} = f_\text{L}\cdot \nprompt_\text{C} = f_\text{S}\cdot \nprompt_\text{B}\, .
\end{equation*}

The yields in the various regions are summarized in \Tab{\ref{tab:bkgdSR}}.  
The vertices in all CRs are used to verify that the designation of one of the $\MuSA$ tracks in the vertex as \MSID\ or \MSonly\ is independent of the designation of the other as \MSID\ or \MSonly\ (the measured correlation is negligibly small, $<0.0015$).  The predicted number of vertices from prompt background processes in the low- and high-mass SRs are $0.14\pm 0.22$ and $0.504 \pm 0.070$, respectively, where the uncertainties are statistical only.
\begin {table}[!htb]
\centering
\caption{
   \label{tab:bkgdSR}
The number of opposite-charge vertices, $\nos$, the number of same-charge vertices, $\nss$, the estimated ratio of opposite-charge to same-charge non-prompt dimuon vertices, $\Rq$ and the estimated number of prompt dimuon vertices $\nprompt$ in each of the control regions B, C and D.  The values of $\nprompt$ are obtained by subtracting the product of $\nss$ and $\Rq$ from $\nos$. The quoted uncertainties in $\nprompt$ include the statistical component and the systematic uncertainty in $\Rq$.}
\begin{tabular}{ l  r  r c  S[table-format=5.0(6)]}
\toprule
Region & $\nos$ & $\nss$ & $\Rq$ & $\nprompt$  \\
\midrule
\multicolumn{5}{c}{Low-mass region} \\
\midrule
B & $124$ & $63$ & $1.28\pm 0.28$ & 43 \pm 23 \\
C & $451$ & $335$ & $1.20\pm 0.20$ & \asymUnc{~~~~~~~49}{74}{49} \\[+2pt]
D & 19599 & $3220$ & $1.20\pm 0.20$ & 15700 \pm 660 \\
\midrule
\multicolumn{5}{c}{High-mass region} \\
\midrule
B & $120$ & $0$ & $1.28\pm 0.28$ & 120 \pm 11 \\
C & $92$ & $0$ & $1.20\pm 0.20$ & 92 \pm 10 \\
D & 21940 & $24$ & $1.20\pm 0.20$ &  21900 \pm 150 \\
\bottomrule
\end{tabular}
\end{table}

As a cross-check, the B, C and D control regions are subdivided into bins of either muon $\pt$ or muon $\eta$ and $\phi$, and the transfer factors and predictions of $\nprompt$ in the SRs are recomputed.  For both the low-mass and high-mass selection, the sum over the predicted prompt background yields in each bin is consistent with the nominal value.

\subsection{Total background}
\label{sec:totalbkgd}

The predicted number of non-prompt muon vertices is summed with the predicted number of prompt muon vertices from SM background processes to give the predicted total number of background vertices in each of the SRs: $13.8 \pm 4.9$ and $0.50^{+1.41}_{-0.07}$ for $\srzd$ and $\srggm$, respectively, where the uncertainties include the statistical components and the systematic uncertainty in $\Rq$
\begin{table}[htb!]
\centering
\caption{
   \label{tab:VR}
Predicted non-prompt $\nfake_\text{VR}$, prompt $\nprompt_\text{VR}$, and total $\nbkgd_\text{VR}$, background yields and number of observed vertices $\nobs_\text{VR}$ in data in the high-mass validation region.  The uncertainty in $\nprompt_\text{VR}$ includes the statistical component and the systematic uncertainty in $\Rq$.}
\begin{tabular}{ l  c }
\toprule
Yield & High-mass validation region \\
\midrule
$\nfake_\text{VR}$ & \asymUnc{~~~~~~~~~~2.5}{2.3}{1.6} \\[+4pt]
$\nprompt_\text{VR}$ & $7.20\pm 0.25$ \\[+2pt]
\midrule
$\nbkgd_\text{VR}$ & \asymUnc{9.7}{2.3}{1.6} \\[+2pt]
\midrule
\midrule
$\nobs_\text{VR}$ & 13 \\
\bottomrule
\end{tabular}
\end{table}

The reliability of the background estimation method is validated by applying it to both the sum of the simulated background samples and to a high-mass validation region in the data.  The predicted number of dimuon vertices in  the simulated sample agrees with the number of observed vertices, to within the statistical precision, in both the low- and high-mass signal regions.  As the simulated samples do not include multijet processes or cosmic muon backgrounds, this is primarily a validation of the technique to estimate the background from prompt dimuon vertices.
The validation region in data comprises dimuon vertices that satisfy all of the selection criteria of the high-mass region, with the exception that the requirement on the transverse boost of the dimuon system is inverted:  it is required to have a value less than two, which ensures that there is negligible contribution from signal processes.  The results are given in \Tab{\ref{tab:VR}}.
These two studies validate the method within the statistical precision.

\section{Systematic uncertainties}
\label{sec:syst}

The systematic uncertainties are described in detail below. They include those in:  the integrated luminosity, used in converting signal yields to cross-sections; the background estimate, derived entirely from the data; and the signal efficiency, determined from MC simulations.  All systematic uncertainties are treated as uncorrelated.

The uncertainty in the 2016 integrated luminosity is $2.2\%$. It is derived, following a methodology similar to that detailed in Ref.~\cite{DAPR-2013-01}, and using the LUCID-2 detector for the baseline luminosity measurements \cite{LUCID2}, from calibration of the luminosity scale using $x$--$y$ beam-separation scans.

Sources of systematic uncertainties in the signal efficiencies include possible mismodeling of the trigger and \MuSA\ efficiencies and pileup effects in the MC simulation.  For the high-mass SR, the uncertainty associated with trigger and \MuSA\ track reconstruction efficiency is determined by comparing the observed yields in the data with MC simulation of $\Zboson +\text{jets}$ events, using the selection criteria of the OS B, C and D control regions and the additional requirement $70 < \mvtx < 110~\gev$.  The difference between the yields in data and the simulated background samples is used to assign a systematic uncertainty of $1\%$ to the combined trigger and \MuSA\  track-reconstruction efficiency.  For the low-mass SR, the efficiency of the trigger and \MuSA\ track reconstruction is compared between MC simulation and data for $J/\psi \to \mu\mu$ events, using a tag-and-probe technique.  The efficiency is measured as a function of the angular separation between the two muons, and a maximum deviation of $6\%$ is observed. This difference is taken as an uncertainty in the signal efficiency.  The agreement between data and MC simulation for the reconstruction efficiency for \MuSA\ tracks with large impact parameters was studied by comparing a cosmic-ray muon simulation to cosmic-ray muon candidates in data~\cite{SUSY-2014-02}.  Comparing the ratio of the muon candidate $d_0$ distributions in the two samples yields a $d_0$-dependent efficiency correction that is between $1\%$ and $2.5\%$, with an average value of $1.5\%$.  The systematic uncertainty on \MuSA\ track reconstruction associated with this procedure is taken from the statistical uncertainty, and is $2\%$ per track in the vertices.

The systematic uncertainty from pileup effects is determined by varying the pileup reweighting of simulated signal events in a manner that spans the expected uncertainty. This results in a systematic uncertainty of $0.2\%$ in the signal efficiency.

The methods used to estimate the background are entirely data-driven, with statistical uncertainties arising from the numbers of events in the CRs.  The non-prompt-vertex background estimate for both signal regions has a systematic uncertainty of $19\%$ associated with knowledge of the charge correlation $\Rq$, as described in \Sect{\ref{sec:fakebkgd}}.  Systematic uncertainties in the estimate of the prompt background are determined by varying the quantity that distinguishes \MSID\ from \MSonly\ muons, the angular distance between the \MuSA\ track and nearest combined-muon track, by $\pm 50\%$ and repeating the ABCD technique described in \Sect{\ref{sec:smbkgd}}.  A $9\%$ difference in the prompt background estimate is observed, and this is taken as a systematic uncertainty.  

\section{Results}
\label{sec:result}

\begin{figure}[!htbp]
\centering
\subfloat[]{\includegraphics[width = 0.49 \textwidth]{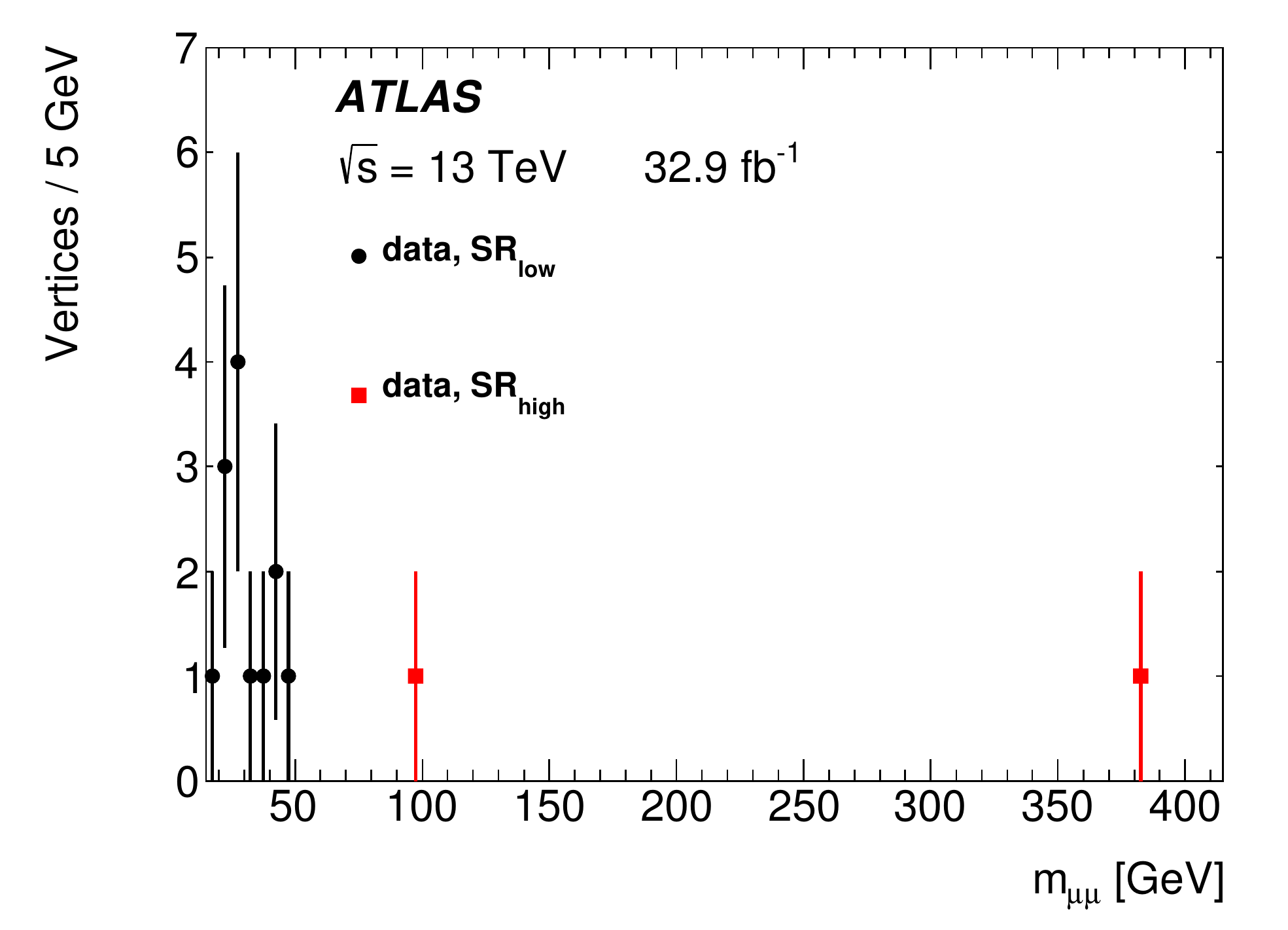}}
\subfloat[]{\includegraphics[width = 0.49 \textwidth]{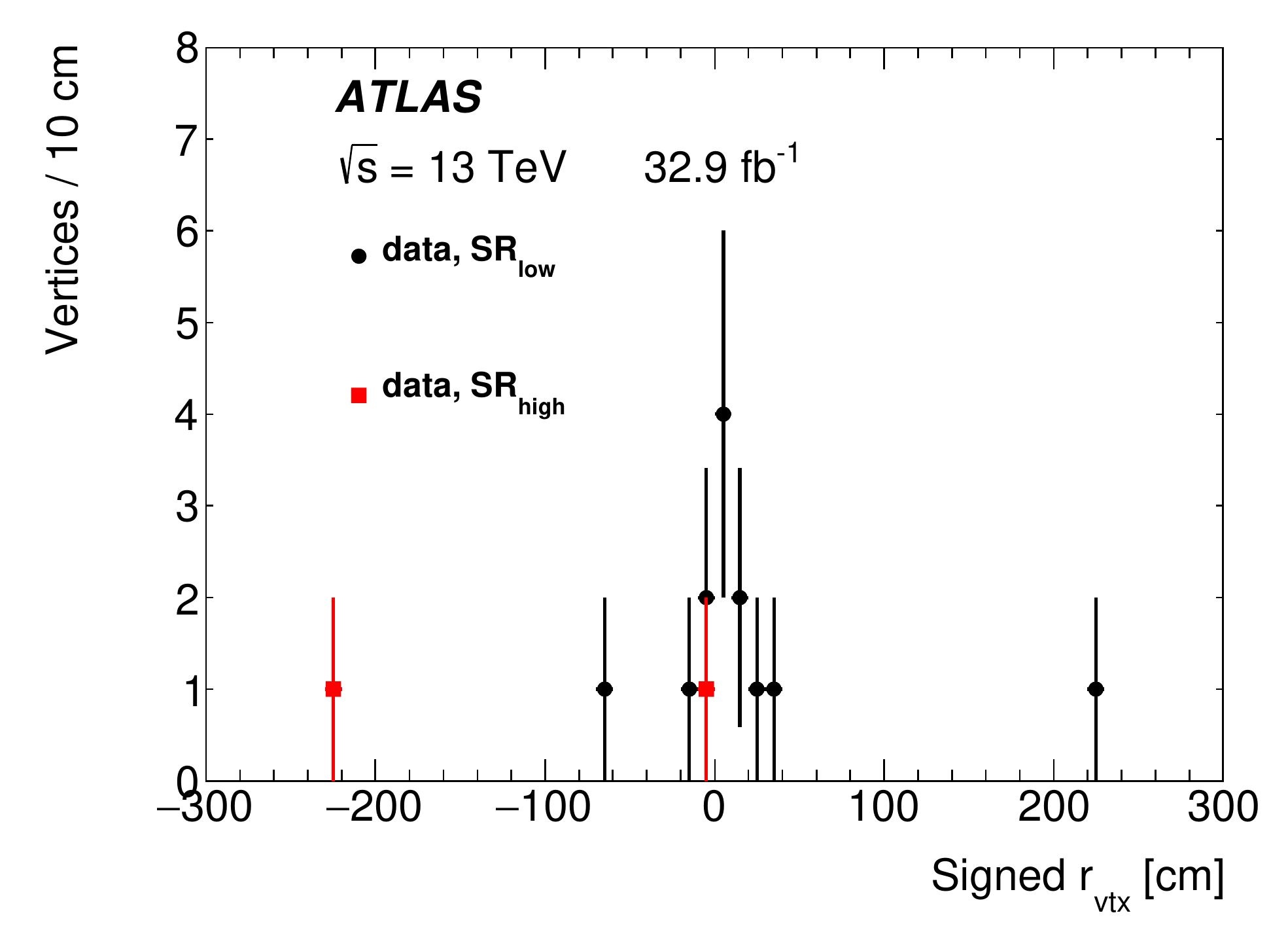}}
\caption{
	\label{fig:Obs}
Distributions of (a) dimuon invariant mass  $\mvtx$ and (b) vertex radius $\rvtx$ for displaced dimuon vertices in the low-mass (black circles) and high-mass (red squares) signal regions.
}
\end{figure}

The predicted number of non-prompt muon vertices is summed with the predicted number of prompt muon vertices from SM background processes to give the predicted total number of background vertices, $\nbkgd$, in each SR.  The predicted background yields, along with the number of observed vertices in the data, are summarized in \Tab{\ref{tab:totalPred}}.  The distributions of $\mvtx$ and $\rvtx$ are shown in \Fig{\ref{fig:Obs}} for the observed vertices in the two signal regions.  Each dimuon vertex is in a separate event, and therefore the number of events observed is equivalent to the number of vertices.  The dimuon vertex with the highest mass has $\mvtx = 381~\gev$, $\rvtx = -220~\cm$ and $\zvtx = 99~\cm$.  Close inspection of the event reveals characteristics of being cosmic in origin.  The observation of one such dimuon vertex in $\srggm$ is consistent, within the uncertainties, with the non-prompt background estimate of $\nfake = 0.0^{+1.4}_{-0.0}$. The other vertex in $\srggm$ has a mass compatible with the decay of the SM $\Zboson$.
The dimuon vertex with the largest value of $\rvtx$ is in $\srzd$ and has $\mvtx = 46~\gev$, $\rvtx = 223~\cm$ and $\zvtx = 56~\cm$.  The vertex is formed by an MS track passing through the top of the detector combined with another MS track passing through the bottom of the detector, with an angle of nearly 180 degrees between them.  This vertex is likely a cosmic-ray muon that narrowly survived the cosmic-ray veto criteria described in \Sect{\ref{sec:vertexing}}.
\begin {table}[htb!]
\centering
\caption{
   \label{tab:totalPred}
Predicted non-prompt $\nfake$, prompt $\nprompt$, and total $\nbkgd$ background yields and number of observed vertices $\nobs$ in data in $\srzd$ and $\srggm$.  The uncertainties in the predicted background yields are statistical uncertainties and systematic uncertainties added in quadrature.}
\begin{tabular}{ l  S[table-format=3.1(3)] c }
\toprule
Yield & $\srzd$ & $\srggm$ \\
\midrule
$\nfake$ & 13.6 \pm 4.9 & \asymUnc{0.0}{1.4}{0.0} \\[+4pt]
$\nprompt$ & \asymUnc{0.1}{0.2}{0.1} & $0.50\pm 0.07$ \\[+2pt]
\midrule
$\nbkgd$ & 13.8 \pm 4.9 & \asymUnc{0.50}{1.42}{0.07} \\[+2pt]
\midrule
\midrule
$\nobs$ & 15 & 2 \\
\bottomrule
\end{tabular}
\end{table}

\begin{figure}[!htbp]
\centering
\subfloat[]{\includegraphics[width = 0.49 \textwidth]{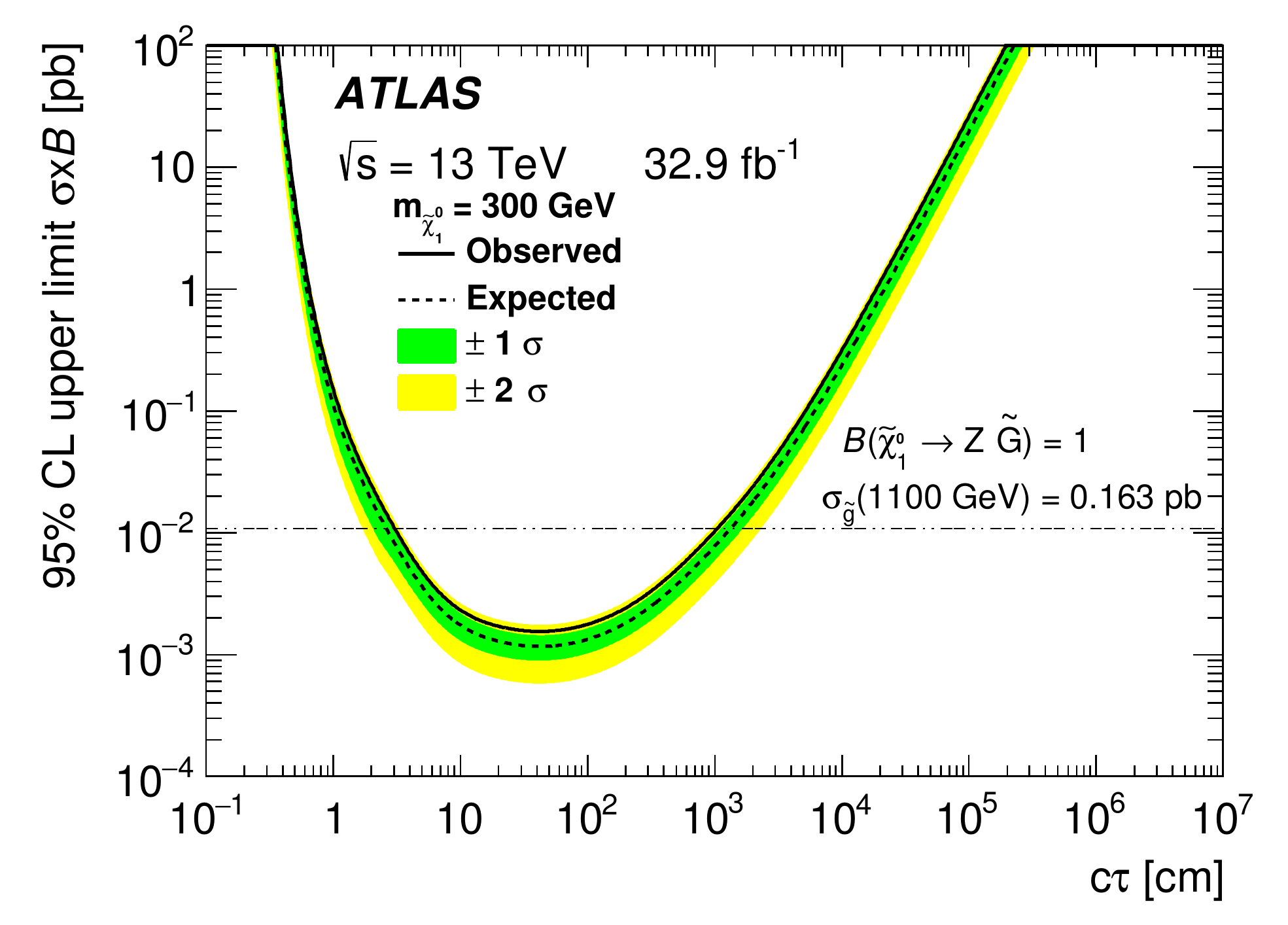}}
\subfloat[]{\includegraphics[width = 0.49 \textwidth]{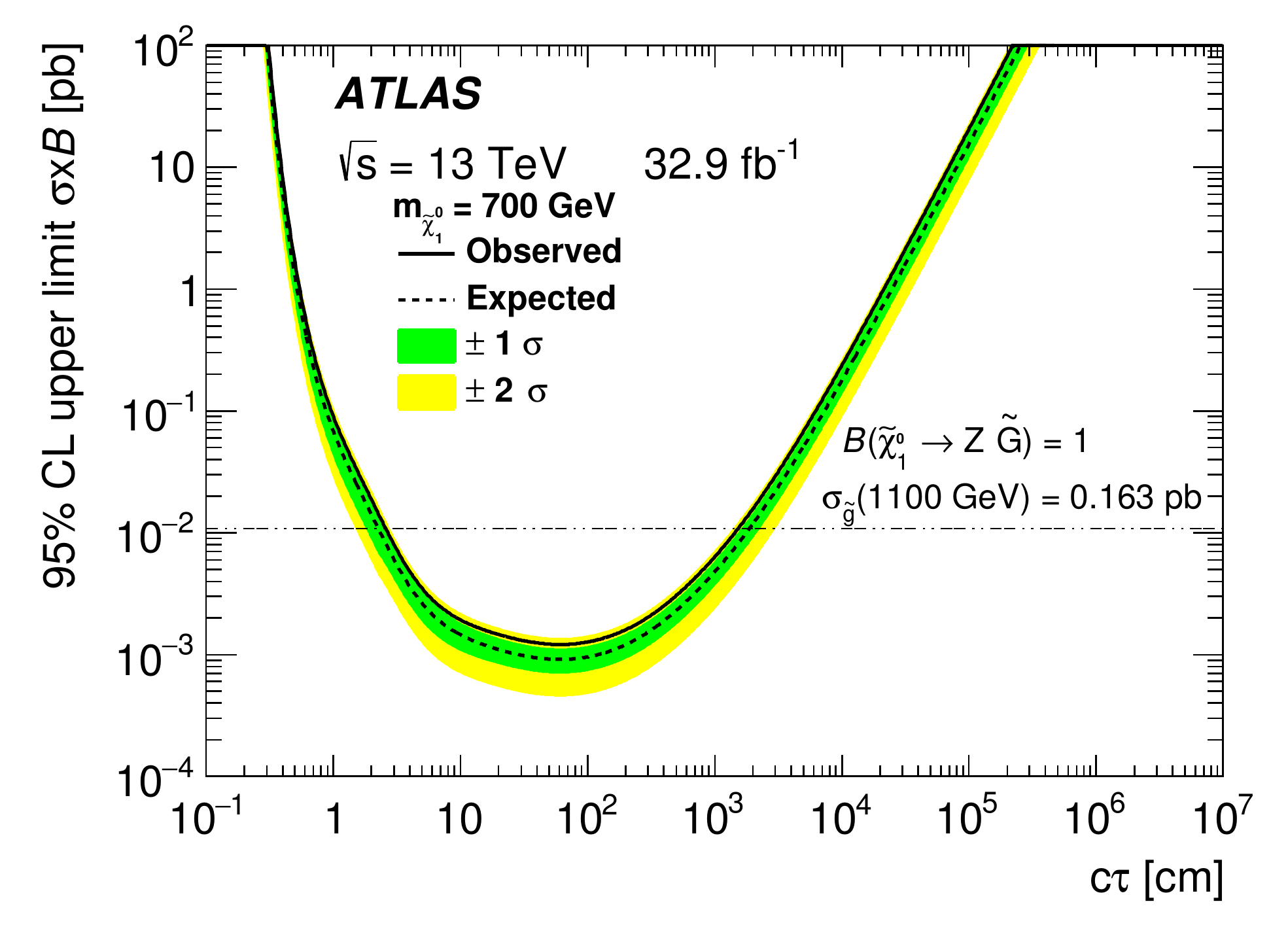}}\\
\subfloat[]{\includegraphics[width = 0.49 \textwidth]{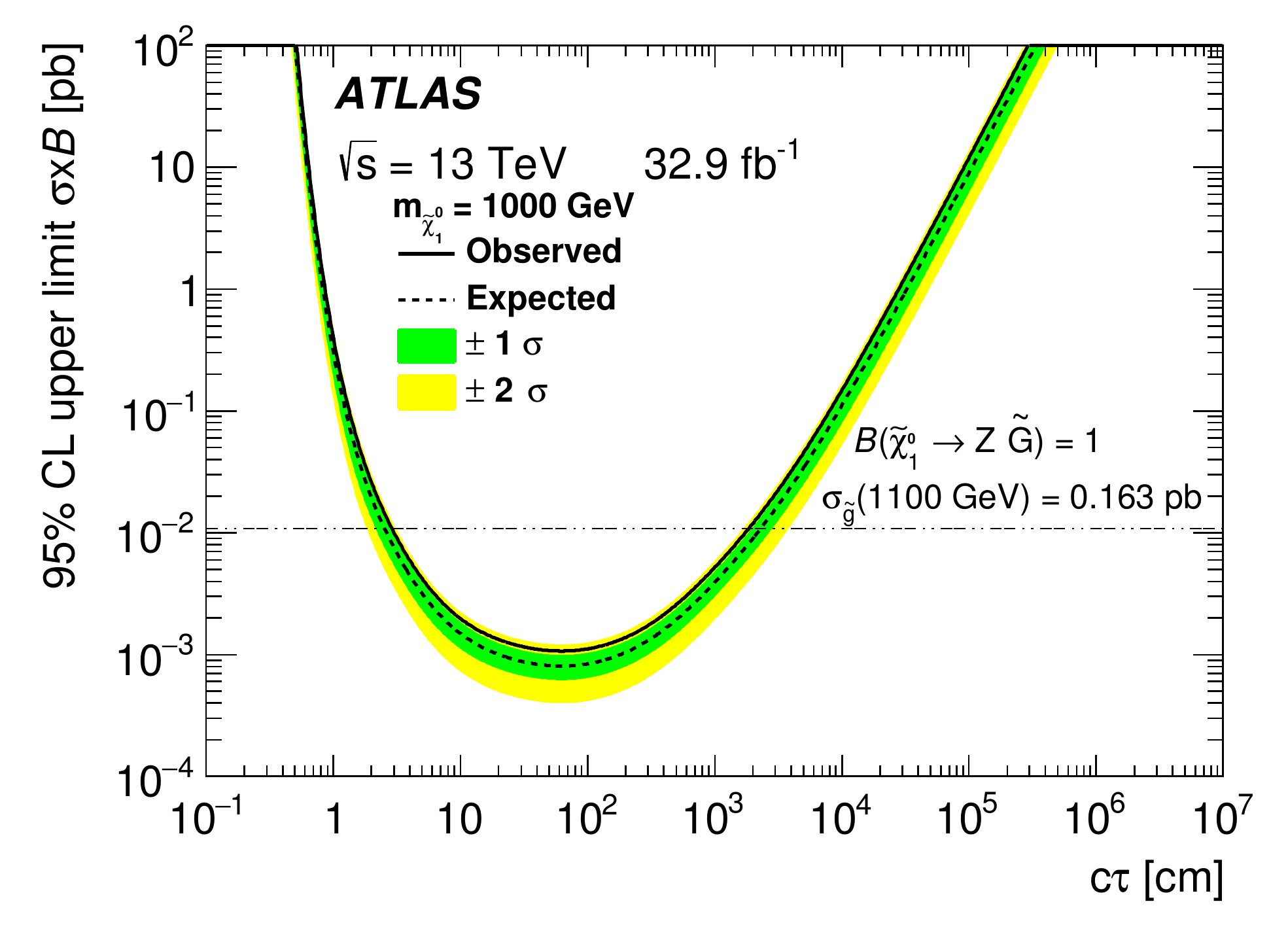}}
\caption{
	\label{fig:ctULGGM}
The observed and expected $95\%$ CL upper limits on the product of cross-section and branching ratios
for pair production of gluinos, leading to a final state of $\mu^+\mu^- + X$, 
in the GGM model, 
as a function of the $\ninoone$ lifetime, for $m_{\gluino}=1100~\gev$ and three different choices of $\mchi$:  (a) $300~\gev$, (b) $700~\gev$ and (c) $1000~\gev$.  The shaded green (yellow) bands represent the $1\sigma$ ($2\sigma$) uncertainties in the expected limits.  The dashed horizontal line represents the value of the cross-section times branching fractions predicted from simulation, with $m_{\gluino}=1100~\gev$, $\sigma\left(pp\to \gluino\gluino \right) = 0.1635~\pb$, $B\left(\ninoone \to \Zboson \gravino \right) = 1.0$ and $B\left(\Zboson \to \mu^+\mu^- \right) = 0.03366$.}
\end{figure}

\begin{figure}[!htbp]
\centering
\subfloat[]{\includegraphics[width = 0.49 \textwidth]{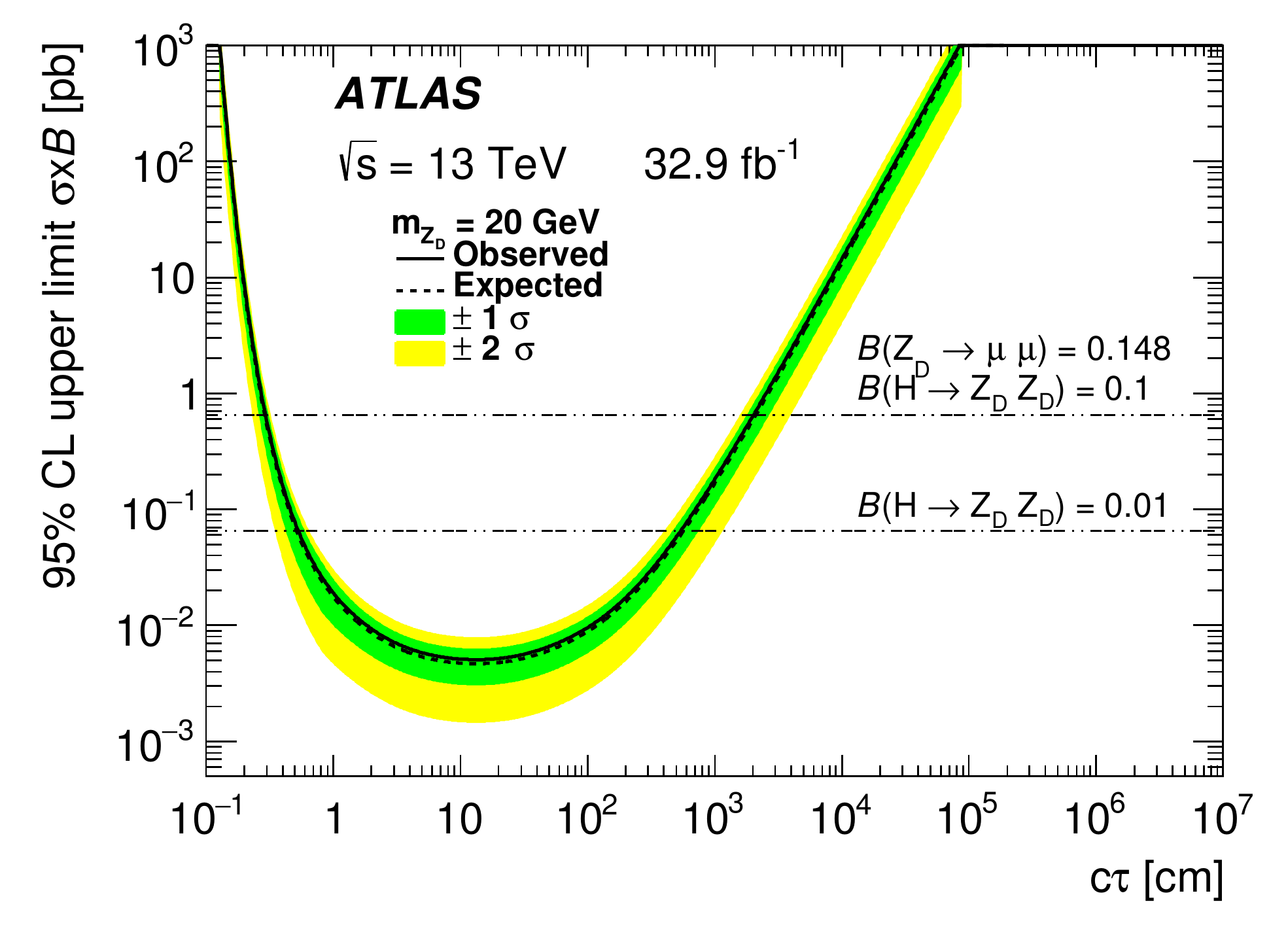}}
\subfloat[]{\includegraphics[width = 0.49 \textwidth]{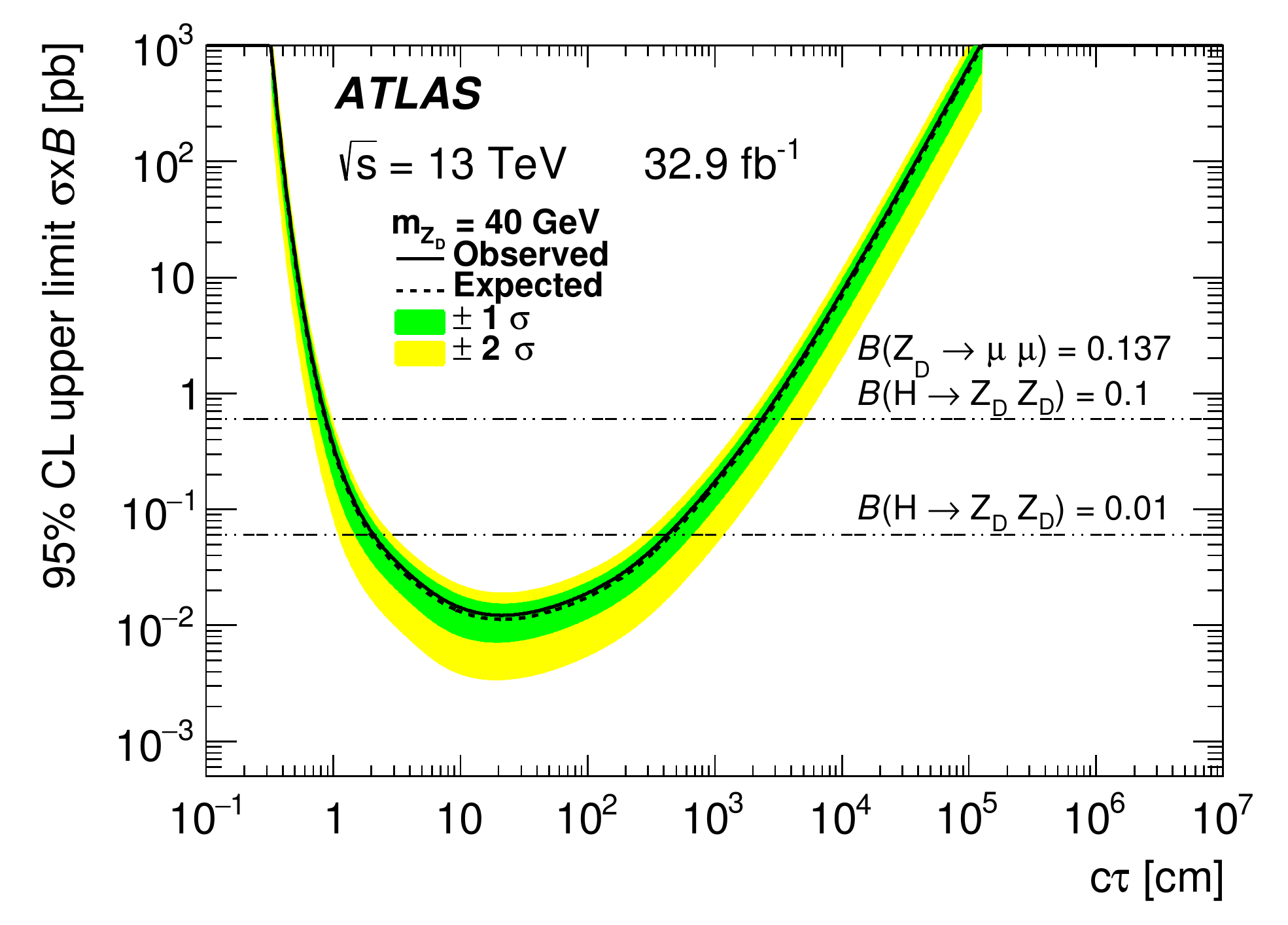}} \\
\subfloat[]{\includegraphics[width = 0.49 \textwidth]{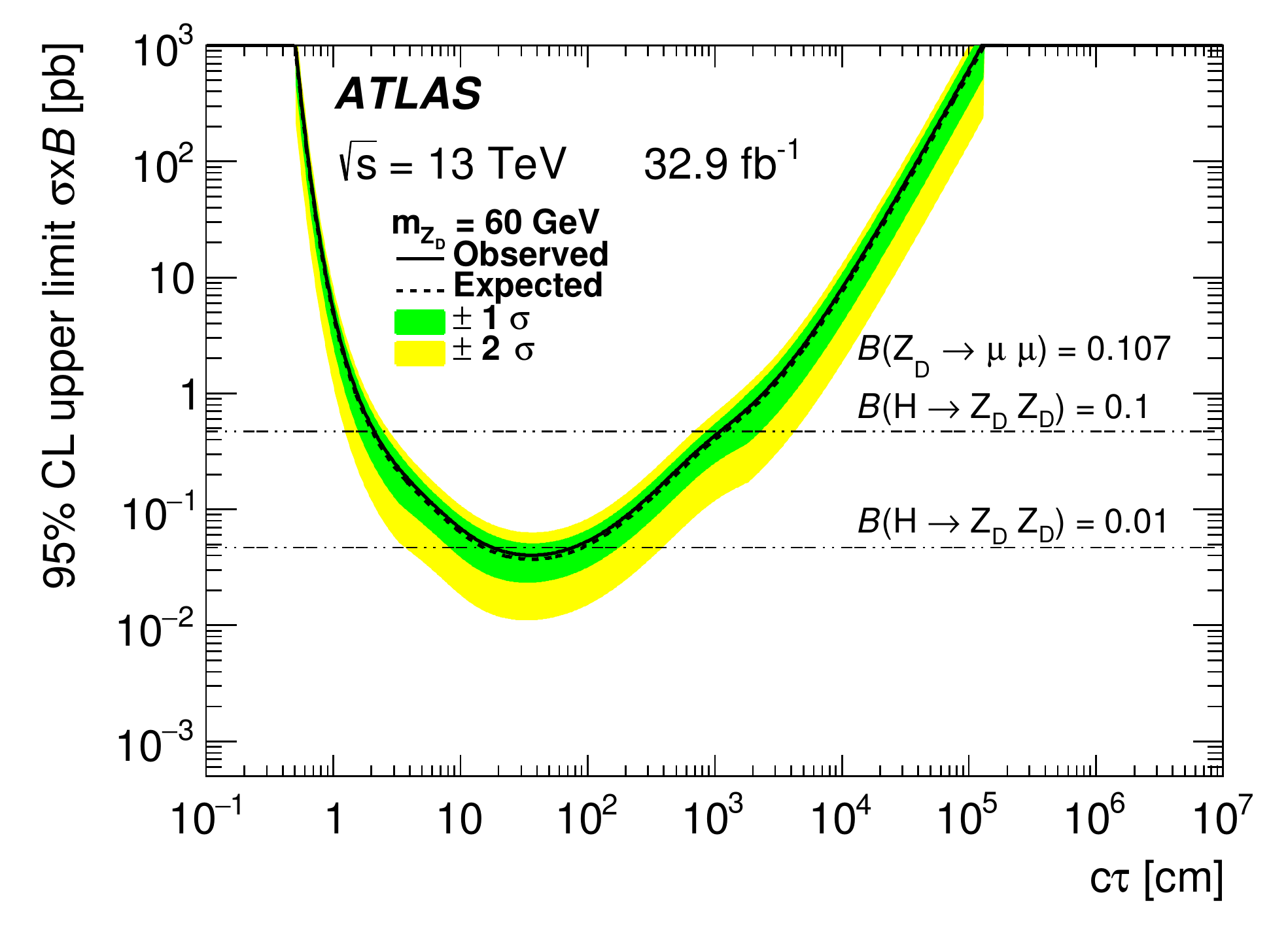}}
\caption{
	\label{fig:ctULZd}
The observed and expected $95\%$ CL upper limits on the product of cross-section and branching ratios, $\sigma\times B = \sigma\left( pp\to \Hboson \right)\times B\left(\Hboson \to \zd \zd\right)\times B\left(\zd \to \mu^+\mu^- \right)$, 
in the dark-sector model, as a function of the $\zd$ lifetime, for three different choices of $\mzd$:  (a) $20~\gev$, (b) $40~\gev$ and (c) $60~\gev$.  The shaded green (yellow) bands represent the $1\sigma$ ($2\sigma$) uncertainties in the expected limits.  The dashed horizontal lines represent the values of the cross-section times branching fractions predicted by simulation, with $m_{\Hboson}=125~\gev$, $m_{\Hd}=300~\gev$, $\sigma\left(pp\to \Hboson \right) = 44.1~\pb$ and assuming $B\left( \Hboson \to \zd\zd \right) = 10\%$ or $1\%$.  The value of $B\left(\zd\to \mu^+\mu^-\right)$ varies between $0.1475$ and $0.1066$ for the range $\mzd = 20$--$60~\gev$.}
\end{figure}

\begin{figure}[!htbp]
\centering
{\includegraphics[width = \textwidth]{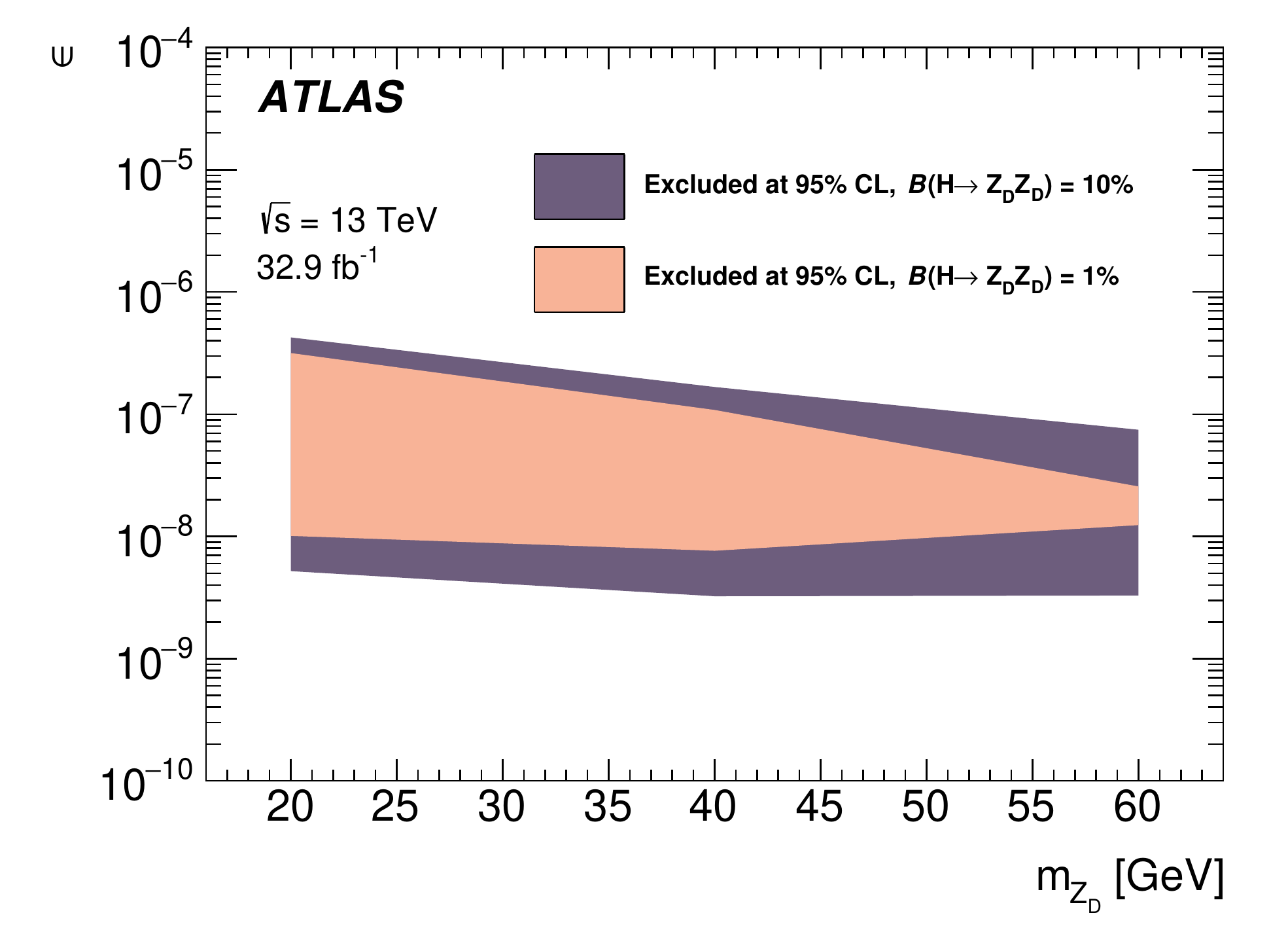}} 
\caption{
\label{fig:epsULZd}
The observed $95\%$ CL excluded regions in the plane of $\zd$--$\Zboson$ kinetic mixing parameter, $\epsilon$, versus $\zd$ mass, for values of $B\left(\Hboson \to \zd \zd \right)=1\%$ or $10\%$, and $m_{\Hd} = 300~\gev$.  The value of $B\left(\zd\to \mu^+\mu^-\right)$ varies between $0.1475$ and $0.1066$ for the range $\mzd = 20$--$60~\gev$.}
\end{figure}

As no significant excess of vertices over the SM background expectation is observed, $95\%$ confidence-level (CL) upper limits on the signal event yields and production cross-sections are calculated for various values of the proper decay distance $c\tau$ of the long-lived particle in each of the two BSM scenarios considered.\footnote{For events that are selected exclusively by the trimuon trigger the observed signal yield will have a quadratic dependence on $B\left(\zd\to\mu^+\mu^-\right)$.  The collimated-dimuon trigger efficiency dominates over the trimuon trigger efficiency for the values of $\mzd$ considered in this paper.}
The limits are calculated using the $\cls$ prescription~\cite{Read:2002hq} with a Poisson likelihood used as the test statistic.  Uncertainties in the signal efficiency and background expectation are included as nuisance parameters, and the $\cls$ values are calculated by generating ensembles of pseudo-experiments corresponding to the background-only and signal-plus-background hypotheses.  Both the expected and observed limits are shown in~\Fig{\ref{fig:ctULGGM}} for the GGM model, and in~\Fig{\ref{fig:ctULZd}} for the dark-sector model, where $\srggm$ is used for the GGM model and $\srzd$ is used for the dark-sector model.  In the GGM model with a gluino mass of $1100~\gev$ and $\ninoone$ masses of $300$, $700$ and $1000~\gev$, $\ctchi$ values are excluded in the ranges $3.1$--$1000~\cm$, $2.6$--$1500~\cm$ and $2.9$--$1800~\cm$, respectively.  The observed limits are about $1.5\sigma$ weaker than the expected limits because of the small excess of events observed in $\srggm$.  In the dark-sector model with a dark-Higgs-boson mass of $300~\gev$, $B\left( \Hboson \to \zd\zd \right) = 10\%$ and $\zd$ masses of $20$, $40$ and $60~\gev$, $\ctzd$ values are excluded in the ranges $0.3$--$2000~\cm$, $0.9$--$2400~\cm$ and $2.1$--$1100~\cm$, respectively.  These limits are translated into $95\%$ exclusion contours in the plane of the $\zd$--$\Zboson$ kinetic mixing parameter, $\epsilon$, and the $\zd$ mass, and are shown in \Fig{\ref{fig:epsULZd}}.  Values of $\epsilon$ of the order $10^{-8}$ are excluded, for $20 < \mzd < 60~\gev$.  

\FloatBarrier

\section{Conclusion}
\label{sec:conclusion}

This article reports on a search for BSM long-lived particles decaying into two muons of opposite-sign electric charge in a sample of $pp$ collisions recorded by the ATLAS detector at the LHC with a center-of-mass energy of $\sqrt{s}=13~\tev$ and an integrated luminosity of $\lumi$.  The search is performed by identifying dimuon vertices with displacements from the $pp$ interaction point in the range of $1$--$400~\cm$ and having invariant mass $\mvtx$ within one of two signal regions:  $20$--$60~\gev$ or $> 60~\gev$.  In neither signal region is a significant excess observed in the number of vertices relative to the predicted background. Hence upper limits at $95\%$ confidence level on the product of cross-section and branching fraction are calculated, as a function of lifetime, for production of long-lived particles in either a dark-sector model with dark-photon masses in the range $20$--$60~\gev$, produced from decays of the Higgs boson, or in a general gauge-mediated supersymmetric model with a gluino mass of $1100~\gev$ and neutralino masses in the range $300$--$1000~\gev$.  For the models considered, the lower and upper lifetime limits are set from $1$ to $2400~\cm$ in c$\tau$, respectively, depending on the targeted model's parameters.

\section*{Acknowledgments}

We thank CERN for the very successful operation of the LHC, as well as the
support staff from our institutions without whom ATLAS could not be
operated efficiently.

We acknowledge the support of ANPCyT, Argentina; YerPhI, Armenia; ARC, Australia; BMWFW and FWF, Austria; ANAS, Azerbaijan; SSTC, Belarus; CNPq and FAPESP, Brazil; NSERC, NRC and CFI, Canada; CERN; CONICYT, Chile; CAS, MOST and NSFC, China; COLCIENCIAS, Colombia; MSMT CR, MPO CR and VSC CR, Czech Republic; DNRF and DNSRC, Denmark; IN2P3-CNRS, CEA-DRF/IRFU, France; SRNSFG, Georgia; BMBF, HGF, and MPG, Germany; GSRT, Greece; RGC, Hong Kong SAR, China; ISF, I-CORE and Benoziyo Center, Israel; INFN, Italy; MEXT and JSPS, Japan; CNRST, Morocco; NWO, Netherlands; RCN, Norway; MNiSW and NCN, Poland; FCT, Portugal; MNE/IFA, Romania; MES of Russia and NRC KI, Russian Federation; JINR; MESTD, Serbia; MSSR, Slovakia; ARRS and MIZ\v{S}, Slovenia; DST/NRF, South Africa; MINECO, Spain; SRC and Wallenberg Foundation, Sweden; SERI, SNSF and Cantons of Bern and Geneva, Switzerland; MOST, Taiwan; TAEK, Turkey; STFC, United Kingdom; DOE and NSF, United States of America. In addition, individual groups and members have received support from BCKDF, the Canada Council, CANARIE, CRC, Compute Canada, FQRNT, and the Ontario Innovation Trust, Canada; EPLANET, ERC, ERDF, FP7, Horizon 2020 and Marie Sk{\l}odowska-Curie Actions, European Union; Investissements d'Avenir Labex and Idex, ANR, R{\'e}gion Auvergne and Fondation Partager le Savoir, France; DFG and AvH Foundation, Germany; Herakleitos, Thales and Aristeia programmes co-financed by EU-ESF and the Greek NSRF; BSF, GIF and Minerva, Israel; BRF, Norway; CERCA Programme Generalitat de Catalunya, Generalitat Valenciana, Spain; the Royal Society and Leverhulme Trust, United Kingdom.

The crucial computing support from all WLCG partners is acknowledged gratefully, in particular from CERN, the ATLAS Tier-1 facilities at TRIUMF (Canada), NDGF (Denmark, Norway, Sweden), CC-IN2P3 (France), KIT/GridKA (Germany), INFN-CNAF (Italy), NL-T1 (Netherlands), PIC (Spain), ASGC (Taiwan), RAL (UK) and BNL (USA), the Tier-2 facilities worldwide and large non-WLCG resource providers. Major contributors of computing resources are listed in Ref.~\cite{ATL-GEN-PUB-2016-002}.

\printbibliography

\clearpage 
 
\begin{flushleft}
{\Large The ATLAS Collaboration}

\bigskip

M.~Aaboud$^\textrm{\scriptsize 34d}$,    
G.~Aad$^\textrm{\scriptsize 99}$,    
B.~Abbott$^\textrm{\scriptsize 124}$,    
O.~Abdinov$^\textrm{\scriptsize 13,*}$,    
B.~Abeloos$^\textrm{\scriptsize 128}$,    
D.K.~Abhayasinghe$^\textrm{\scriptsize 91}$,    
S.H.~Abidi$^\textrm{\scriptsize 164}$,    
O.S.~AbouZeid$^\textrm{\scriptsize 143}$,    
N.L.~Abraham$^\textrm{\scriptsize 153}$,    
H.~Abramowicz$^\textrm{\scriptsize 158}$
% ...
\bigskip
\\

% ...
$^{13}$Institute of Physics, Azerbaijan Academy of Sciences, Baku; Azerbaijan.\\
% ...
$^{34}$$^{(a)}$Facult\'e des Sciences Ain Chock, R\'eseau Universitaire de Physique des Hautes Energies - Universit\'e Hassan II, Casablanca;$^{(b)}$Centre National de l'Energie des Sciences Techniques Nucleaires (CNESTEN), Rabat;$^{(c)}$Facult\'e des Sciences Semlalia, Universit\'e Cadi Ayyad, LPHEA-Marrakech;$^{(d)}$Facult\'e des Sciences, Universit\'e Mohamed Premier and LPTPM, Oujda;$^{(e)}$Facult\'e des sciences, Universit\'e Mohammed V, Rabat; Morocco.\\
% ...
$^{91}$Department of Physics, Royal Holloway University of London, Egham; United Kingdom.\\
% ...
$^{99}$CPPM, Aix-Marseille Universit\'e, CNRS/IN2P3, Marseille; France.\\
% ...
$^{124}$Homer L. Dodge Department of Physics and Astronomy, University of Oklahoma, Norman OK; United States of America.\\
% ...
$^{128}$LAL, Universit\'e Paris-Sud, CNRS/IN2P3, Universit\'e Paris-Saclay, Orsay; France.\\
% ...
$^{143}$Santa Cruz Institute for Particle Physics, University of California Santa Cruz, Santa Cruz CA; United States of America.\\
% ...
$^{153}$Department of Physics and Astronomy, University of Sussex, Brighton; United Kingdom.\\
% ...
$^{158}$Raymond and Beverly Sackler School of Physics and Astronomy, Tel Aviv University, Tel Aviv; Israel.\\
% ...
$^{164}$Department of Physics, University of Toronto, Toronto ON; Canada.\\
% ...
$^{a}$ Also at  Department of Physics, University of Malaya, Kuala Lumpur; Malaysia.\\
$^{b}$ Also at Borough of Manhattan Community College, City University of New York, NY; United States of America.\\
$^{c}$ Also at Centre for High Performance Computing, CSIR Campus, Rosebank, Cape Town; South Africa.\\
$^{d}$ Also at CERN, Geneva; Switzerland.\\
$^{e}$ Also at CPPM, Aix-Marseille Universit\'e, CNRS/IN2P3, Marseille; France.\\
% ...
$^{*}$ Deceased

\end{flushleft}


\end{document}